%% file: MICCAI_2020.tex
\let\origvec\vec 
\documentclass[runningheads]{llncs}

\let\vec\origvec
\usepackage{graphicx}
\usepackage{amsmath}
\usepackage{bm}                                     
\usepackage{amsfonts}                               
\usepackage{xcolor}
\usepackage[toc, acronym, nopostdot]{glossaries}    
\usepackage[ruled,vlined]{algorithm2e}
\usepackage[utf8]{inputenc}
\usepackage{hyperref}
\usepackage[capitalise]{cleveref}                               
\usepackage[caption=false]{subfig}                  
\usepackage{lipsum}
\usepackage{booktabs}       

\newcommand{\fnurl}[1]{\footnote{\href{#1}{\texttt{#1}}}}

\input{Acronyms}

\begin{document}
\title{Simulation of Brain Resection for Cavity Segmentation Using Self-Supervised and Semi-Supervised Learning}  
\titlerunning{Simulation of Brain Resection for Cavity Segmentation}
%
\author{
    Fernando Pérez-García\inst{1,2}\orcidID{0000-0001-9090-3024}  
    \and Roman Rodionov \inst{3,4}                                
    \and Ali Alim-Marvasti \inst{1,3,4}                           
    \and Rachel Sparks \inst{2}                                   
    \and John S. Duncan \inst{3,4}                                
    \and Sébastien Ourselin \inst{2}                              
}
%
\authorrunning{F. Pérez-García et al.}
%
\institute{
    Wellcome EPSRC Centre for Interventional and Surgical Sciences (WEISS), University College London, London, UK
    \and School of Biomedical Engineering and Imaging Sciences (BMEIS), King's College London, London, UK
    \and Department of Clinical and Experimental Epilepsy, UCL Queen Square Institute of Neurology, London, UK
    \and National Hospital for Neurology and Neurosurgery, Queen Square, London, UK
    \email{fernando.perezgarcia.17@ucl.ac.uk}
}
\maketitle              
%

\glsunset{mri}

\begin{abstract}
\input{sections/abstract}

\glsresetall

\keywords{%
    Neurosurgery
    \and Segmentation
    \and Self-supervised learning
}
\end{abstract}

\glsunset{mri}

\input{sections/introduction}
\input{sections/methods}

\input{sections/results}
\input{sections/discussion}
\input{sections/acknowledgments}

%
%
%

\bibliographystyle{splncs04}
\bibliography{MICCAI_2020}

\clearpage

\input{sections/supplementary}

\end{document}

%% file: Acronyms.tex
\newacronym{adni}{ADNI}{Alzheimer's Disease Neuroimaging Initiative}
\newacronym{amp}{AMP}{automatic mixed precision}
\newacronym{cnn}{CNN}{convolutional neural network}
\newacronym{csf}{CSF}{cerebrospinal fluid}
\newacronym{dsc}{DSC}{Dice score}
\newacronym{ez}{EZ}{epileptogenic zone}
\newacronym{gif}{GIF}{geodesical information flows}
\newacronym{ixi}{IXI}{Information eXtraction from Images}
\newacronym{mri}{MRI}{magnetic resonance imaging}
\newacronym{nhnn}{NHNN}{National Hospital for Neurology and Neurosurgery}
\newacronym{nhs}{NHS}{National Health Service}
\newacronym{oasis}{OASIS}{Open Access Series of Imaging Studies}
\newacronym{prelu}{PReLU}{Parametric Rectified Linear Unit}

%% file: sections/abstract.tex

Resective surgery may be curative for drug-resistant focal epilepsy,
but only 40\% to 70\% of patients achieve seizure freedom after surgery.
Retrospective quantitative analysis could elucidate patterns
in resected structures and patient outcomes to improve resective surgery.
However, the resection cavity must first be segmented
on the postoperative MR image.
\Glspl{cnn} are the state-of-the-art image segmentation technique,
but require large amounts of annotated data for training.
Annotation of medical images is a time-consuming process requiring
highly-trained raters, and often suffering from high inter-rater
variability.
Self-supervised learning can be used to generate training instances
from unlabeled data.
%
%
%
We developed an algorithm to simulate resections on preoperative MR images.
We curated a new dataset, EPISURG, comprising 431 postoperative and 269 preoperative MR images
from 431 patients who underwent resective surgery.
In addition to EPISURG, we used three public datasets comprising 1813 preoperative MR images for training.
%
%
We trained a 3D \gls{cnn} on artificially resected images created on the fly during training, using images from
1) EPISURG,
2) public datasets and
3) both.
To evaluate trained models, we calculate \gls{dsc} between model segmentations and 200 manual annotations performed by three human raters.
%
%
%
The model trained on data with manual annotations obtained a
median (interquartile range) \gls{dsc} of 65.3 (30.6).
The \gls{dsc} of our best-performing model,
trained with no manual annotations,
is 81.7 (14.2).
For comparison, inter-rater agreement between human annotators was 84.0 (9.9).
%
%
%
We demonstrate a training method for \glspl{cnn} using simulated resection cavities that can accurately segment real resection cavities, without manual annotations.

%% file: sections/introduction.tex
\section{Introduction}


Only 40\% to 70\% of patients with refractory focal epilepsy are seizure-free
after resective surgery~\cite{jobst_resective_2015}.
Retrospective studies relating
clinical features and resected brain structures
(such as amygdala or hippocampus) to surgical outcome may provide
useful insight to identify and guide resection of the epileptogenic zone.
%
To identify resected structures, first,
the resection cavity must be segmented on the postoperative MR image.
Then, a preoperative image with a corresponding brain parcellation
can be registered to the postoperative MR image to identify resected structures.

In the context of brain resection, the cavity fills with
\gls{csf} after surgery~\cite{winterstein_partially_2010}.
This causes an inherent uncertainty in resection cavity delineation
when adjacent to sulci, ventricles, arachnoid cysts or oedemas,
as there is no intensity gradient separating the structures.
Moreover, brain shift can occur during surgery,
causing regions outside the cavity to fill with \gls{csf}.

Decision trees have been used for brain cavity
segmentation from $T_2$-weighted,
FLAIR, and pre- and post-contrast $T_1$-weighted \gls{mri} in the context of
glioblastoma surgery~\cite{meier_automatic_2017,herrmann_fully_2018}.
%
%
%
Relatedly, some methods have simulated or segmented brain lesions
to improve non-linear registration with missing correspondences.
%
%
Brett et al.~\cite{brett_spatial_2001} propagated lesions manually segmented from pathological brain images to structurally normal brain images by registering images to a common template space.
Removing the lesion from consideration when computing the similarity metric improved non-linear registration.
Methods to directly compute missing correspondences during registration,
which can give an estimate of the resection
cavity, have been proposed~\cite{chitphakdithai_non-rigid_2010,chen_deformable_2015,drobny_handling_2015}.
Pezeshk et al.~\cite{pezeshk_seamless_2017} trained a series of machine learning classifiers to detect lesions in chest CT scans.
The dataset was augmented by propagating lesions from pathological lungs to healthy lung tissue, using Poisson blending.
This data augmentation technique improved classification results for all machine learning techniques considered.

In traditional machine learning,
data is represented by hand-crafted features which may not be optimal.
In contrast, deep learning, which has been successfully applied to
brain image segmentation~\cite{kamnitsas_efficient_2017,li_compactness_2017},
implicitly computes a problem-specific feature representation.
However, deep learning techniques rely on large annotated datasets for training.
Annotated medical imaging datasets are often small due to the financial and time
burden annotating the data, and the need for highly-trained raters.
Self-supervised learning generates training instances using unlabeled data
from a source domain to learn features
that can be transferred to a target domain~\cite{jing_self-supervised_2019}.
Semi-supervised learning uses labeled as well as unlabeled data
to train models~\cite{van_engelen_survey_2020}.
These techniques can be used to leverage unlabeled medical imaging data
to improve training in instances where acquiring annotations
is time-consuming or costly.

We present a fully-automatic algorithm to simulate resection cavities from preoperative
$T_1$-weighted MR images,
applied to self-supervised learning for brain resection cavity segmentation.
We validate this approach by comparing models trained with and without
manual annotations, using 200 annotations from three human raters
on 133 postoperative MR images with lobectomy or lesionectomy
(133 annotations to test models performance and 67 annotations to
assess inter-rater variability).


%% file: sections/methods.tex
\section{Methods}

\input{sections/methods_simulation}
\input{sections/methods_data}
\input{sections/methods_implementation}

%% file: sections/methods_simulation.tex
\subsection{Resection Simulation}
\label{sec:simulation}

\newcommand{\p}{\bm{p}}
\newcommand{\vv}{\bm{v}}
\newcommand{\X}{\bm{X}}
\newcommand{\Y}{\bm{Y}}
\newcommand{\M}{\bm{M}}
\newcommand{\R}{\mathbb{R}}

We generate automatically a training instance $(\X_R, \Y_R)$
representing a resected brain $\X_R$ and its corresponding
cavity segmentation $\Y_R$ from a preoperative image~$\X$ using the following approach.

\begin{figure} [b!]
    \centering

    \subfloat[\label{fig:sphere}]{%
        \scalebox{0.75}{\includegraphics[%
            width=0.166666\textwidth,
            trim={540 0 0 0},
            clip
        ]{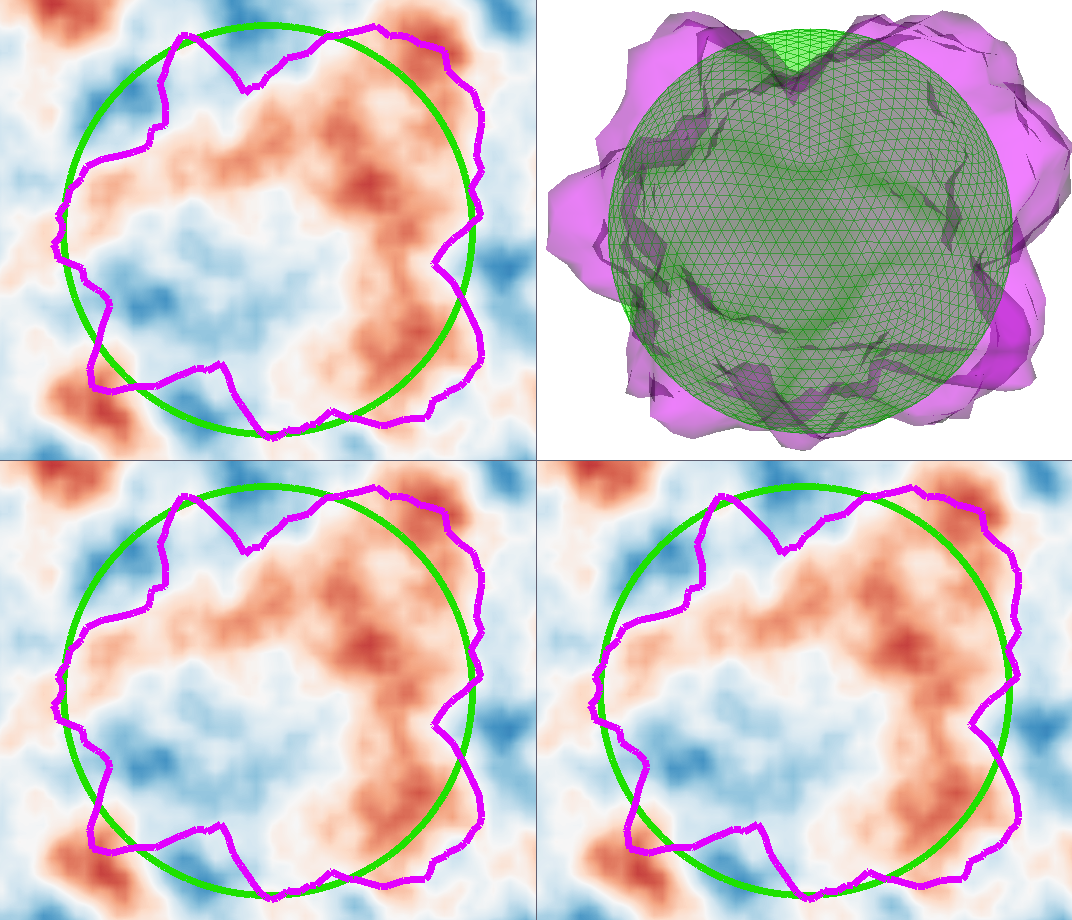}}
    }%
    \subfloat[\label{fig:ellipsoid_p}]{%
        \scalebox{0.98}{\includegraphics[%
            width=0.166666\textwidth,
            trim = {0 50 1236 65},
            clip
        ]{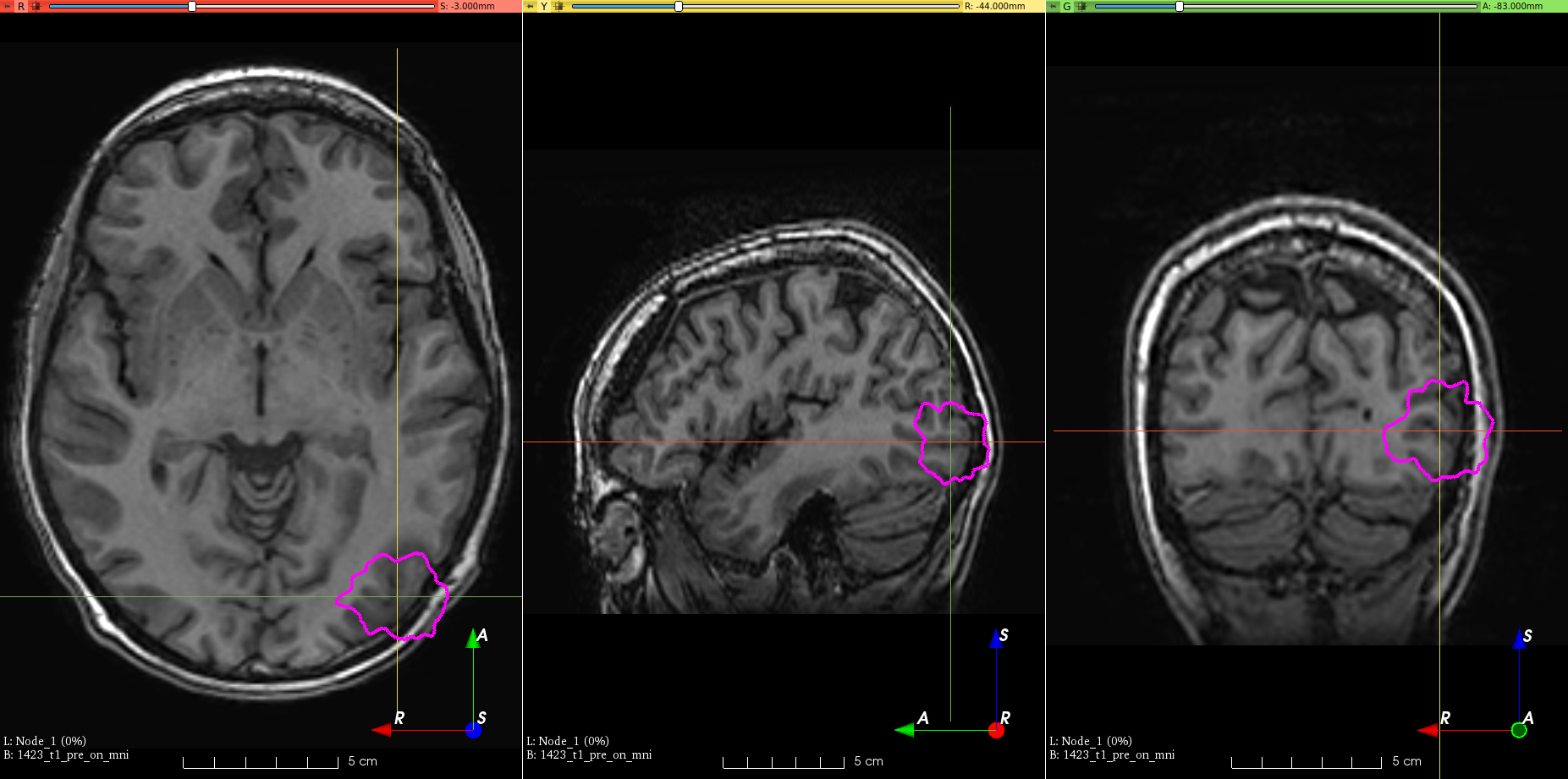}}
    }%
    \subfloat[\label{fig:resectable_1423}]{%
        \scalebox{0.98}{\includegraphics[%
            width=0.166666\textwidth,
            trim = {0 50 1236 65},
            clip
        ]{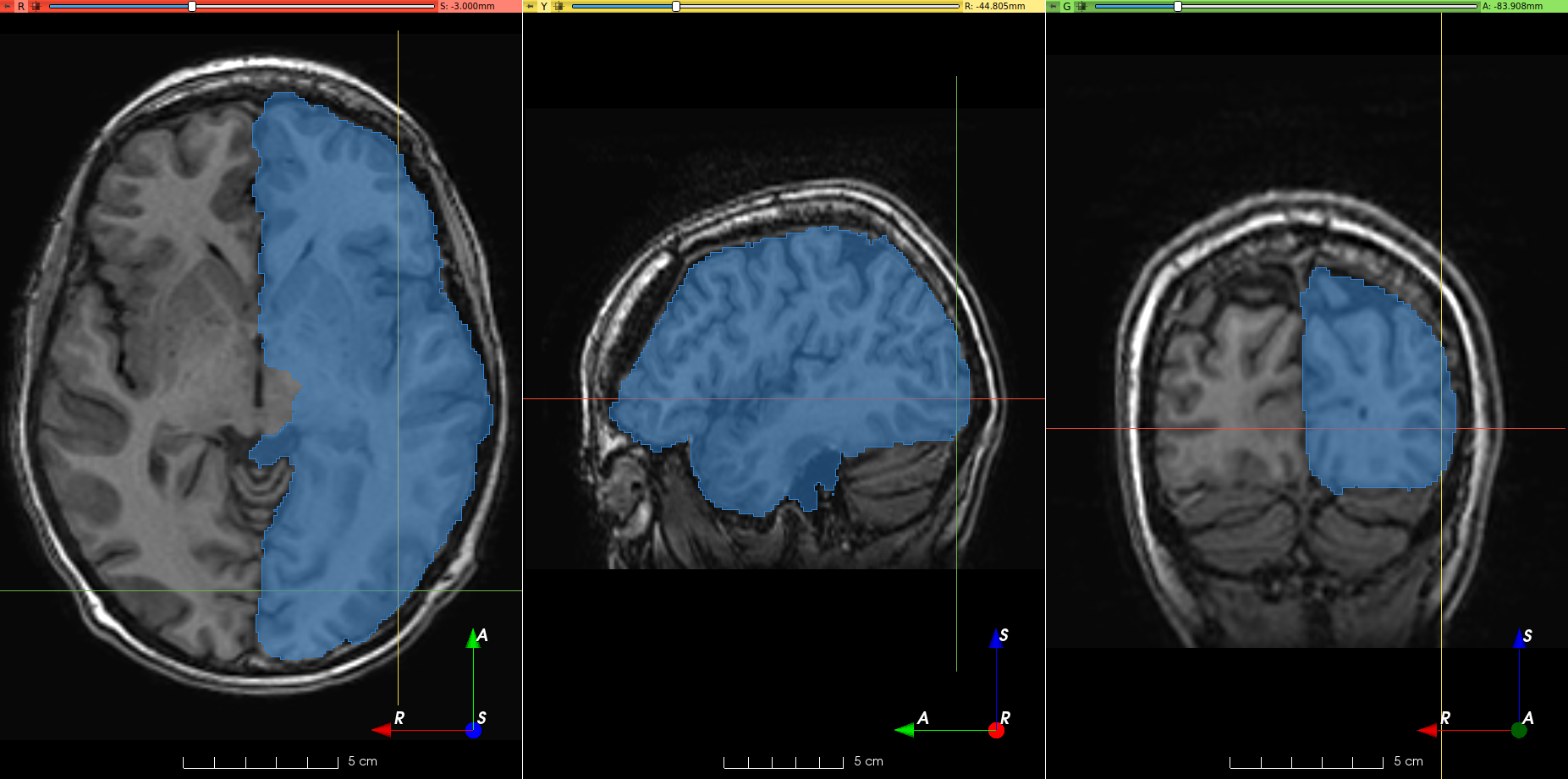}}
    }%
    \subfloat[\label{fig:resection_axial_label}]{%
        \scalebox{0.98}{\includegraphics[%
            width=0.166666\textwidth,
            trim = {0 50 0 65},
            clip
        ]{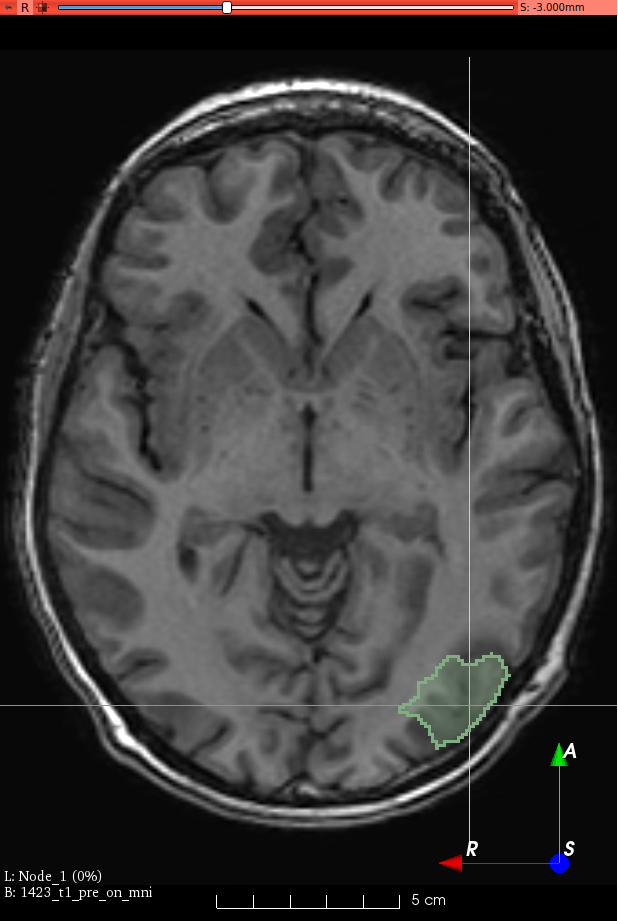}}
    }%
    \subfloat[\label{fig:resection_axial_resected}]{%
        \scalebox{0.98}{\includegraphics[%
            width=0.166666\textwidth,
            trim = {0 50 0 65},
            clip
        ]{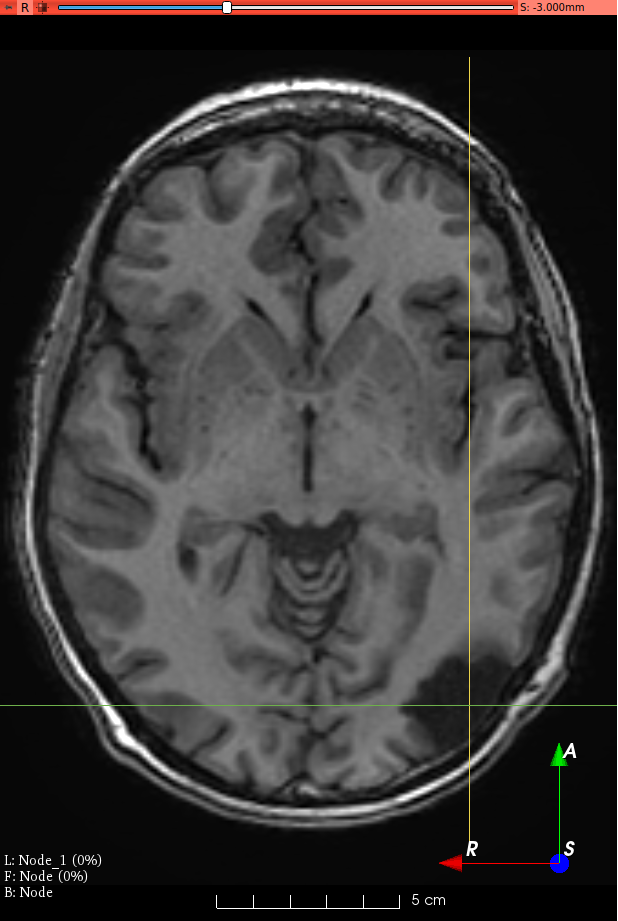}}
    }%
    \subfloat[\label{fig:resection_axial_original}]{%
        \scalebox{0.98}{\includegraphics[%
            width=0.166666\textwidth,
            trim = {0 50 0 65},
            clip
        ]{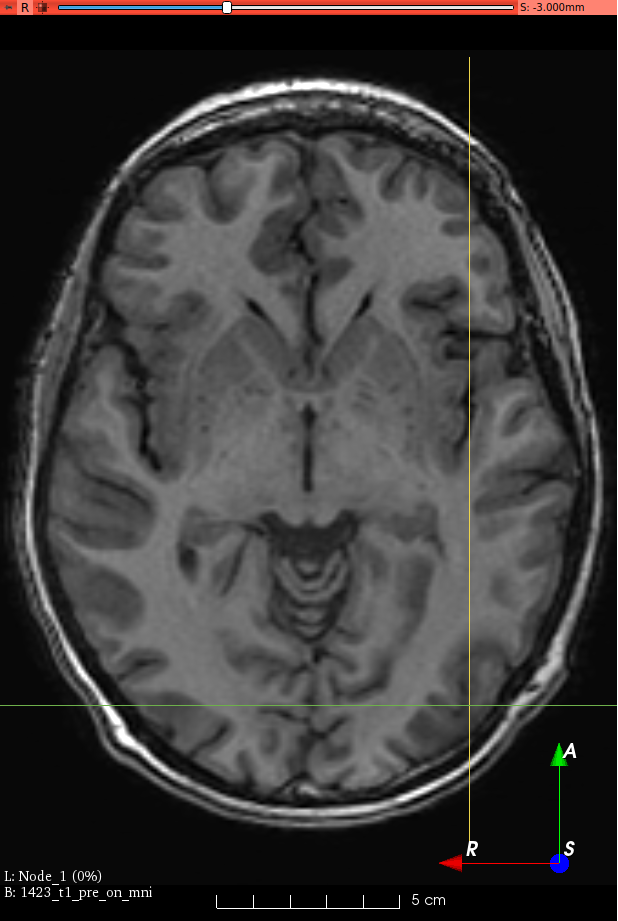}}
    }

    \caption{%
    Resection simulation.
        (a) Sphere surface mesh
        before ($S$, green) and after ($S_\delta$, magenta)
        perturbation.
        $S$ and $S_\delta$ (top);
        intersection of $S$ and $S_\delta$ with a plane of the
        simplex noise volume, generated only for visualization purposes,
        with values between $-1$ (blue) and $1$ (red).
        Radial displacement is proportional to the noise at each vertex~${\vv_i \in V}$
        (b) Transformed mesh $S_E$
        (c) Resectable hemisphere mask $\M_R$
        (d) Simulated resection label $\Y_R$
        (e) Simulated resected image $\X_{R}$
        (f) Original image $\X$.
    }
    \label{fig:resection}
\end{figure}

\subsubsection{Resection Label}

A geodesic polyhedron with frequency $f$
is generated by subdividing the edges of an icosahedron $f$ times
and projecting each vertex onto a parametric sphere with unit radius.
This polyhedron models a spherical surface $S = \{ V, F \}$
with vertices
$
    V = \left\{
        \vv_i \in \R^3
    \right\}
    _{i = 1}^{n_V}
$
and faces
$
    F = \left\{
        \bm{f}_k
    \right\}
    _{k = 1}^{n_F}
$.
Each face $\bm{f}_k = \{ i_1^k, i_2^k, i_3^k \}$ is defined as a sequence of
three non-repeated vertex indices.

$S$ is perturbed with simplex noise~\cite{perlin_improving_2002},
a smooth noise generated by interpolating pseudorandom gradients
defined on a multidimensional simplicial grid.
Simplex noise was selected as it is often used to simulate natural-looking textures or terrains.
The noise at point $\p \in \R^3$
is computed as a weighted sum of the noise contribution of $\omega$ different octaves,
with weights
$
        \gamma ^ {n - 1}
        : n \in \left\{ 1, 2, \dots, \omega \right\}
$
controlled by the persistence parameter $\gamma$.
The displacement $\delta : \R^3 \to [-1, 1]$ is proportional
to the noise function $\phi : \R^3 \to [0, 1]$:
\begin{equation}
    \delta(\p)
    = 2 \phi \left( \frac{\p + \bm{\mu} }{\zeta}, \omega, \gamma \right) - 1
\end{equation}
where $\zeta$ is a scaling parameter to control smoothness
and $\bm{\mu}$ is a shifting parameter that adds stochasticity
(equivalent to a random number generator seed).
Each vertex $\vv_i \in V$ is displaced radially by:
\begin{equation}
    \vv_{\delta i}
    = \vv_i
    + \delta(\vv_i)
    \frac{\overrightarrow{\vv_i}}{\|\overrightarrow{\vv_i}\|},
    \qquad \forall i \in \{1, 2, \dots, n_V\}
\end{equation}
to create a perturbed sphere $S_{\delta} = \{ V_{\delta}, F \}$
with vertices
$
V_{\delta} = \left\{
    \vv_{\delta i}
    \right\}
    _{i = 1}^{n_V}
$ (\cref{fig:sphere}).

A series of transforms is applied to $S_{\delta}$
to modify its volume, shape and position.
Let $T_T(\p)$, $T_S(\bm{s})$ and $T_R(\bm{\theta})$ be translation, scaling and rotation transforms.

Perturbing $\vv_i \in V$ shifts the centroid of $S_{\delta}$ off the origin.
$S_{\delta}$ is recentered at the origin by applying
the translation $T_T(\bm{-c})$ to each vertex, where
${\bm{c} = \frac{1}{n_V} \sum_{i = 1}^{n_V} \vv_{\delta i}}$
is the centroid of $S_{\delta}$.

Random rotations around each axis are applied to $S_{\delta}$ with the rotation matrix
$T_R(\bm{\theta}_r) = R_x(\theta_x) \circ R_y(\theta_y) \circ R_z(\theta_z)$,
where~$\circ$~indicates a transform composition,
$R_i(\theta_i)$ is a rotation of $\theta_i$ radians around axis $i$,
and $\theta_i \sim \mathcal{U}(0, 2 \pi)$.

A scaling transform $T_S(\bm{r})$ is applied to $S_{\delta}$,
where $(r_1, r_2, r_3) = \bm{r}$ are the semi-axes of an ellipsoid
with volume $v$ modeling the cavity shape.
The semi-axes are computed as
$r_1 = r$, $r_2 = \lambda r$ and $r_3 = r /\lambda$,
where $r = (3 v / 4)^{1/3}$ and
$\lambda$ controls the semi-axes length ratios.

$S_{\delta}$ is translated such that it is centered at a voxel in the cortical gray matter as follows.
A $T_1$-weighted MR image is defined as $\bm{I}_{MRI} : \Omega \to \R$, where $\Omega \in \R^3$.
A full brain parcellation~$\bm{G} : \Omega \to Z$ is generated for $\bm{I}_{MRI}$
using \acrlong{gif}~\cite{cardoso_geodesic_2015},
where $Z$ is the set of segmented brain structures.
A cortical gray matter mask $\M_{GM}^h : \Omega \to \{0, 1\}$
of hemisphere $h$ is extracted from $\bm{G}$,
where $h$ is randomly chosen from $H = \{\text{left}, \text{right}\}$ with equal probability.
A random gray matter voxel $\bm{g} \in \Omega$ is selected such that
$\M_{GM_h}(\bm{g}) = 1$.
%

The transforms are composed as
${T_E = T_T(\bm{g}) \circ T_S(\bm{r}) \circ T_R(\bm{\theta}_r) \circ T_T(-\bm{c})}$
and applied to $S_{\delta}$ to obtain the resection surface
$S_E = T_E \circ S_{\delta}$.
A mask
\linebreak
${\M_E : \Omega \to \{0, 1\}}$ is generated
from $S_E$ such that
$\M_E(\p) = 1$ for all $\p$ within the cavity and
$\M_E(\p) = 0$ outside.

If $\M_E$ is used as the final mask, the resection might span both hemispheres
or include non-realistic tissues such as bone or scalp (\cref{fig:ellipsoid_p}).
To eliminate this unrealistic scenario, a `resectable hemisphere mask' is generated from the parcellation
as $\M_R (\p) = 1$ if
${\bm{G}(\p) \neq \{\M_{BG}, \M_B, \M_C, \M_{\hat{H}} \} }$
and $0$ otherwise,
where $\M_{BG}$, $\M_B$, $\M_C$ and $\M_{\hat{H}}$ are the sets of labels in $Z$
corresponding to the background, brainstem, cerebellum and contralateral hemisphere, respectively.
%
%
$\M_R$ is smoothed using a series of binary morphological operations (\cref{fig:resectable_1423}).
%
The final resection label used for training is
$\Y_R(\p) = \M_E(\p) \M_R(\p)$ (\cref{fig:resection_axial_label}).









\subsubsection{Resected Image}

To mimic partial volume effects near cavity boundaries,
a Gaussian filter is applied to $\M_R(\p)$
to smooth the alpha channel ${\bm{A} : \Omega \to [0, 1]}$,
defined as
$
    \bm{A}(\p)
    = \M_R(\p)
    * \bm{G}_{\mathcal{N}}(\bm{\sigma}),
    \forall \p \in \Omega,
$
where
$*$ is the convolution operator
and $\bm{G}_{\mathcal{N}}(\bm{\sigma}_{\bm{A}})$ is a Gaussian kernel with standard deviations
$\bm{\sigma}_{\bm{A}} = (\sigma_x, \sigma_y, \sigma_z)$.

To generate a realistic \acrshort{csf} texture,
we create a ventricle mask
\linebreak
${\M_V : \Omega \to \{ 0, 1 \}}$ from $\bm{G}$, such that
$\M_V(\p) = 1$ for all $\p$ within the ventricles and
$\M_V(\p) = 0$ outside.
%
%
Intensity values within ventricles are assumed to have
a normal distribution~\cite{gudbjartsson_rician_1995}
with a mean $\mu_{CSF}$ and standard deviation $\sigma_{CSF}$
calculated from voxel intensity values in $\bm{I}_{MRI}(\p) : \forall \p \in \Omega, \M_V(\p) = 1$.
%
A \acrshort{csf}-like image ${\bm{I}_{CSF} : \Omega \to \R}$ is then generated as
$
    \bm{I}_{CSF}(\p)
    \sim \mathcal{N}(\mu_{CSF}, \sigma_{CSF}),
    \forall \p \in \Omega
$,
and the resected image (\cref{fig:resection_axial_resected}) is the convex combination:
\begin{equation}
    \X_R(\p)
    = \bm{A}(\p) \bm{I}_{CSF}(\p),
    + \left[ 1 - \bm{A}(\p) \right] \bm{I}_{MRI}(\p)
    \qquad \forall \p \in \Omega
\end{equation}

%% file: sections/methods_data.tex
\subsection{Dataset Description}
\label{sec:data}

$T1$-weighted MR images were collected from publicly available datasets \gls{ixi} (566), \gls{adni}(467), and \gls{oasis} (780), for a total of 1813 images.
EPISURG was obtained from patients with refractory
focal epilepsy who underwent resective surgery
at the \gls{nhnn}, London, United Kingdom.
This was an analysis of anonymized data that had been previously acquired as a
part of clinical care, so individual patient consent was not required.
In total there were 431 patients with postoperative $T_1$-weighted MR images,
269 of which had a corresponding preoperative MR image.
%
All images were registered to a common template space using NiftyReg~\cite{modat_global_2014}.


Three human raters annotated a subset of the postoperative images in
\linebreak
EPISURG.
Rater A segmented the resection cavity in 133 images.
These annotations were used to test the models.
This set was randomly split into 10 subsets, where the distribution
of resection types (e.g.\ temporal, frontal, etc.) in each subset is
similar.
To quantify inter-rater variability, Rater B annotated subsets 1 and 2 (34 images),
and Rater C annotated subsets 1 and 3 (33 images).
%

%% file: sections/methods_implementation.tex
\subsection{Network Architecture and Implementation Details}

We used the PyTorch deep learning framework,
training with automatic mixed precision
on two 32-GB TESLA V100 GPUs.
We implemented a variant of 3D U-Net~\cite{cicek_3d_2016} using
two downsampling and upsampling blocks,
trilinear interpolation for the synthesis path,
and 1/4 of the filters for each convolutional layer.
This results in a model with 100 times fewer parameters
than the original 3D U-Net, reducing overfitting and computational burden.
We used dilated convolutions~\cite{chen_deeplab_2017},
starting with a dilation factor of one, then
increased or decreased in steps of one
after each downsampling or upsampling block, respectively.
Batch normalization and PReLU activation functions
followed each convolutional layer.
Finally, a dropout layer with probability 0.5
was added before the output classifier
.
%
%
We used an Adam optimizer
~\cite{kingma_adam_2014}
with an initial learning rate of~$10^{-3}$
and weight decay of~$10^{-5}$.
Training occurred for 60 epochs,
and the learning rate was divided by 10 every 20 epochs.
%
%
A batch size of 8 (4~per GPU) was used for training.
%
90\% of the images were used for training and 10\% for validation.


We wrote and used TorchIO~\cite{perez-garcia_torchio_2020}
to process volumes on the fly during training.
The preprocessing and random augmentation transforms used were
1) simulated resection (see \cref{sec:simulation}),
2) MRI k-space motion artifact~\cite{shaw_mri_2019},
3) histogram standardization~\cite{nyul_new_2000},
4) MRI bias field artifact~\cite{sudre_longitudinal_2017},
5) normalization to zero-mean and unit variance of the foreground voxels, computed using the intensity mean as a threshold~\cite{nyul_new_2000},
6) Gaussian noise,
7) flipping in the left-right direction,
8)~scaling and rotation,
and 9) B-spline elastic deformation.
The resection simulation was implemented as a TorchIO~\cite{perez-garcia_torchio_2020} transform and the code is available online\fnurl{https://github.com/fepegar/resector}.
%

The following parameters were used to generate simulated resections
(see \cref{sec:simulation}):
$f = 16$,                                           
$\omega = 4$,                                       
$\gamma = 0.5$,                                     
$\zeta = 3$,                                        
$\bm{\mu} \sim \mathcal{U}(0, 1000)$,               
$\lambda \sim \mathcal{U}(1, 2)$,                   
and ${\bm{\sigma}_{\bm{A}} \sim \mathcal{U}(0.5, 1)}$.   
%
The ellipsoid volume $v$ is sampled from volumes
of manually segmented cavities from Rater A (see \cref{sec:data}).

%% file: sections/results.tex
\section{Experiments and Results}

We trained models with seven different dataset configurations to assess
how simulated resection cavities impact model accuracy
a) using datasets of similar size and scanner,
b) using datasets of similar size and different scanner,
c) using much larger datasets ($10 \times$ increase) and
d) combined with semi-supervised learning.

\begin{figure}[b!]
    \includegraphics[%
        width=\textwidth,
        trim={0 10 0 10},
        clip
    ]{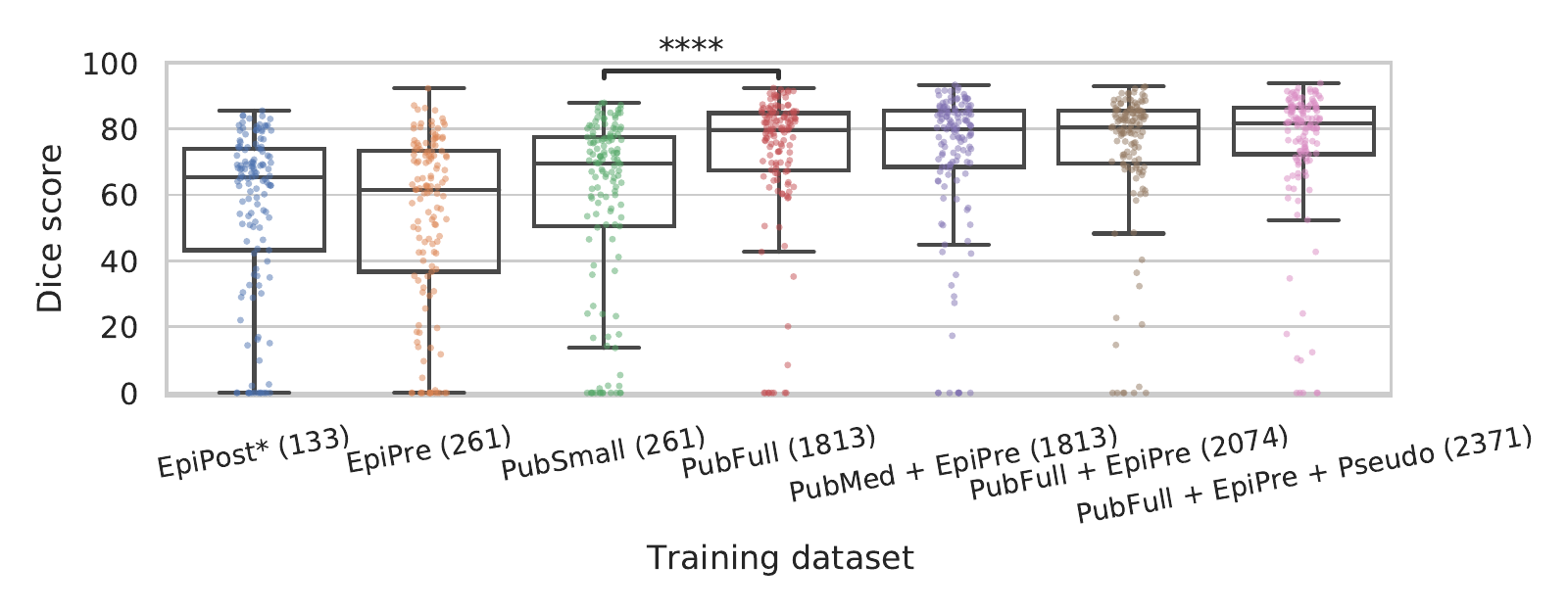}
    \caption{%
        \gls{dsc} values between manual annotations from Rater A
        and segmentations for models.
        Values in brackets indicate number of training subjects.
        Note that only the first model was trained with manual annotations.
        \textit{EpiPost}: postoperative images in EPISURG with manual annotations (the asterisk * indicates fully supervised training with 10-fold cross-validation);
        \textit{EpiPre}: preoperative images from subjects not contained in \textit{EpiPost});
        \textit{PubFull}: public datasets;
        \textit{PubSmall}, \textit{PubMed}: subsets of \textit{PubFull};
        \textit{Pseudo}: pseudo-labeled postoperative images in EPISURG.
    }
    \label{fig:boxes}
\end{figure}

All overlap measurements are expressed as
`median (interquartile range)' \gls{dsc} with respect to the 133 annotations
obtained from Rater~A.
Quantitative results
are shown in \cref{fig:boxes}.

Differences in model performance were analyzed by a
one-tailed
\linebreak
Mann-Whitney~$U$ test with a
significance threshold of $\alpha = 0.05$,
with Bonferroni correction for the seven experiments evaluated
$\left(
    \frac{\alpha}{7 \times (7 - 1)} \approx 0.002
\right)$.

\subsection{Small Datasets}

\begin{figure}[t!]
    \centering

    \subfloat[\label{fig:qualitative_small}]{%
        \scalebox{0.99}{\includegraphics[%
            width=0.155\textwidth,
            trim = {0 0 0 15},
            clip
        ]{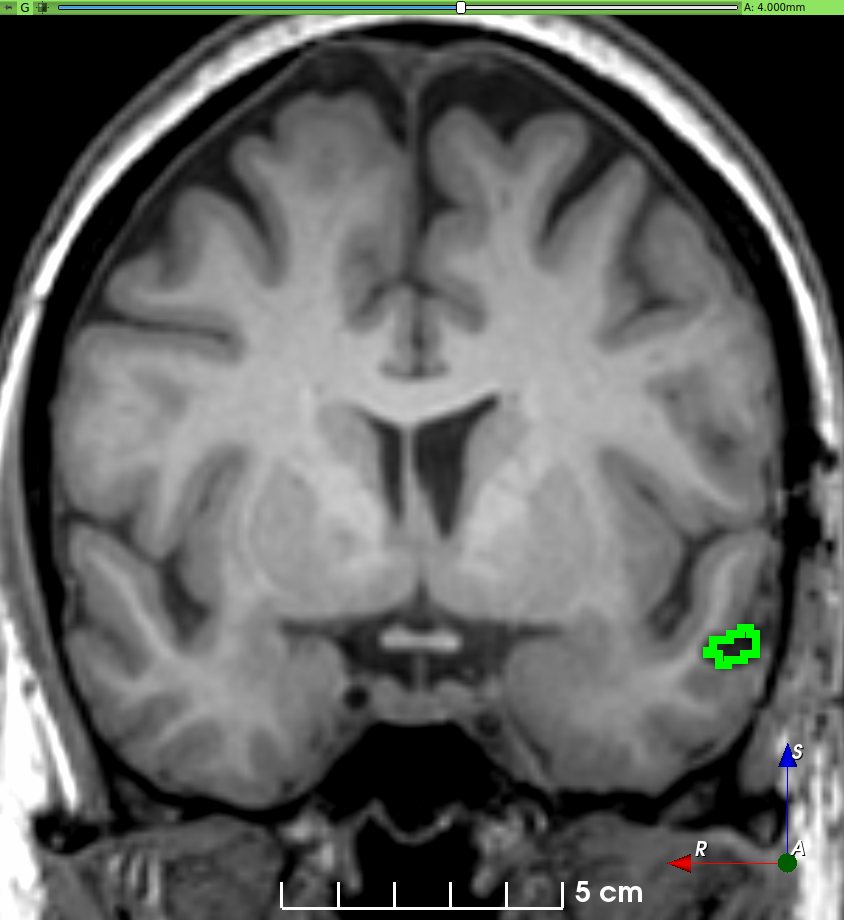}}
    }%
    \subfloat[\label{fig:qualitative_texture}]{%
        \scalebox{0.99}{\includegraphics[%
            width=0.155\textwidth,
            trim = {0 0 0 15},
            clip
        ]{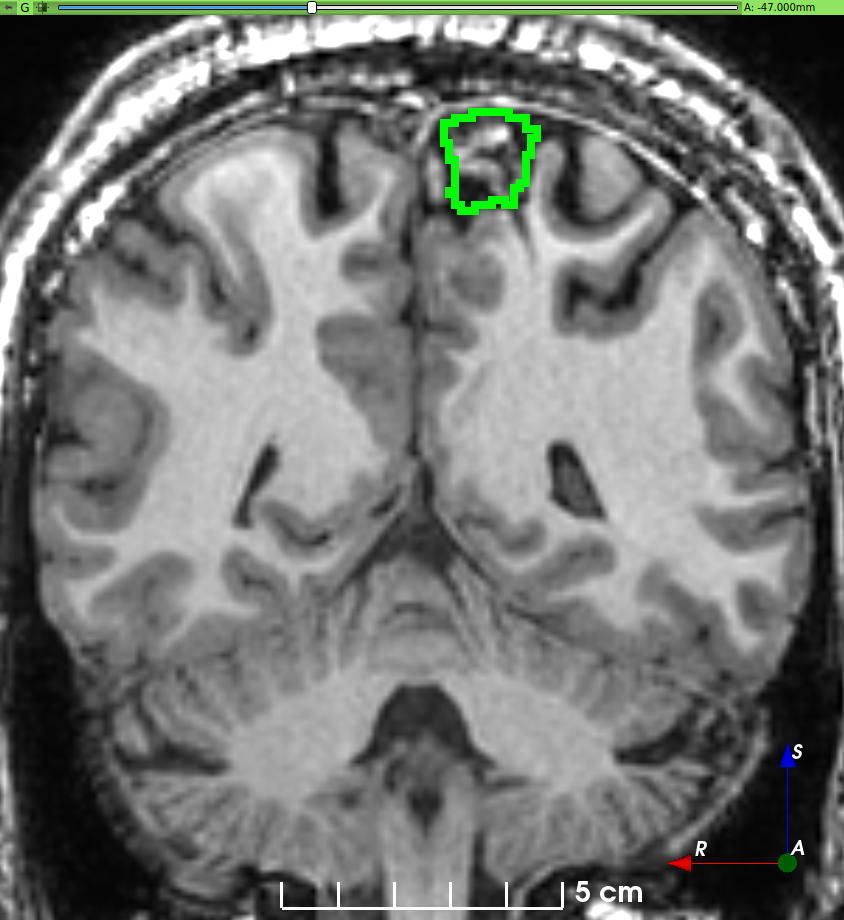}}
    }%
    \subfloat[\label{fig:qualitative_brain_shift}]{%
        \scalebox{0.99}{\includegraphics[%
            width=0.155\textwidth,
            trim = {0 0 0 15},
            clip
        ]{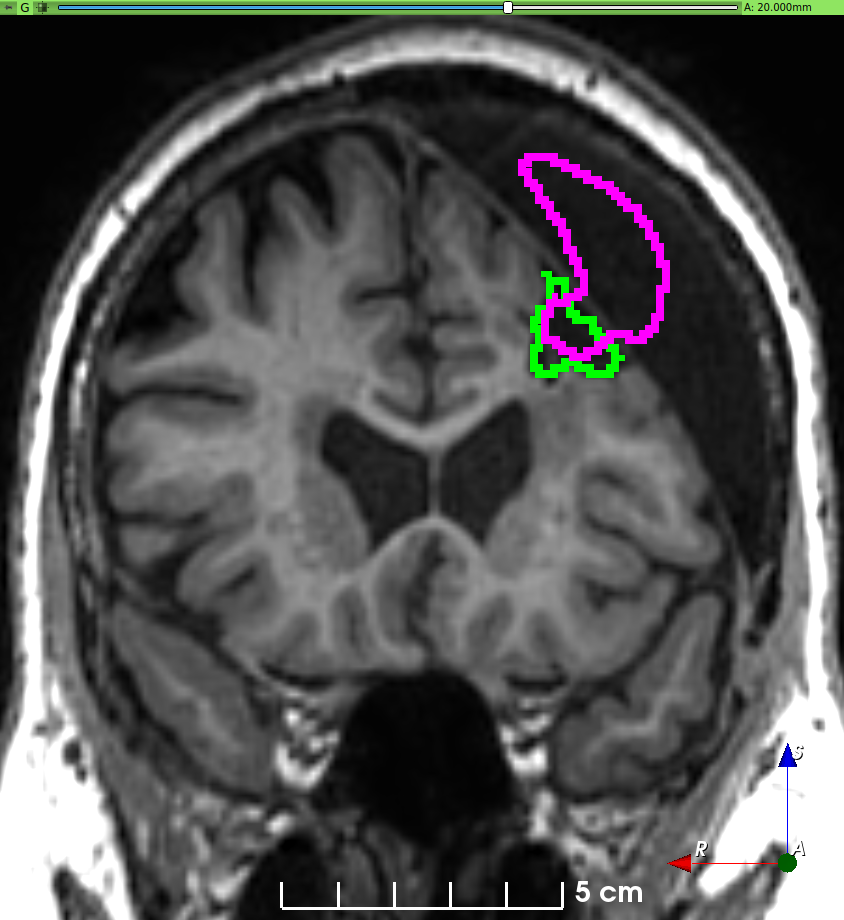}}
    }\hfil%
    \subfloat[\label{fig:qualitative_50}]{%
        \scalebox{0.99}{\includegraphics[%
            width=0.155\textwidth,
            trim = {0 0 0 15},
            clip
        ]{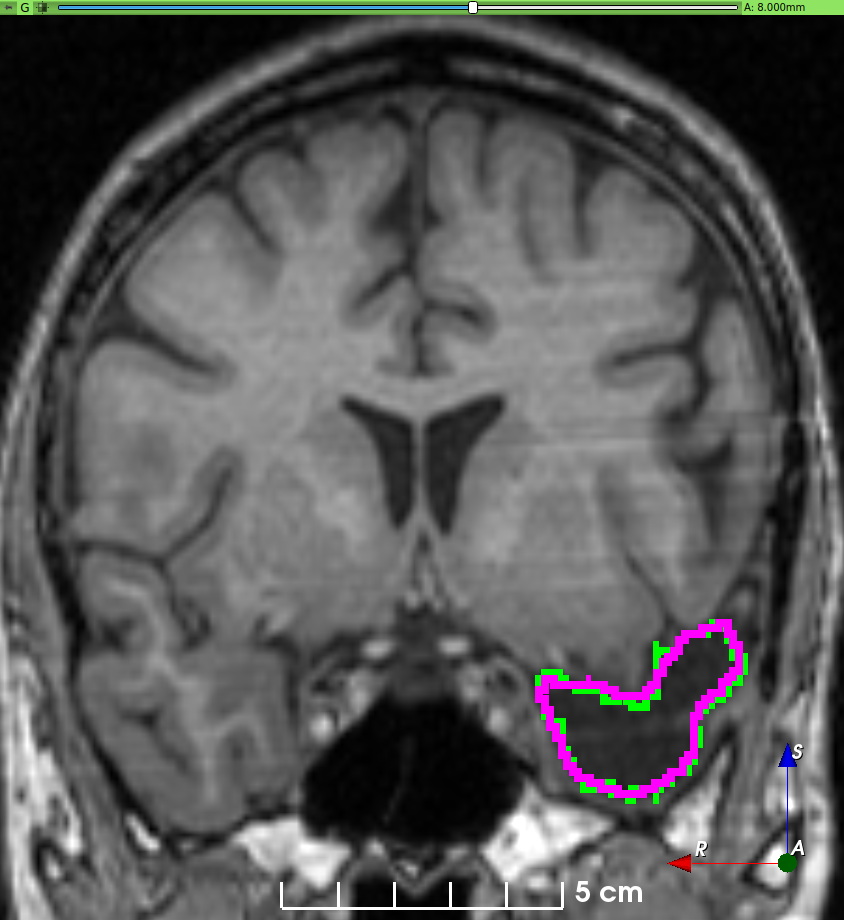}}
    }%
    \subfloat[\label{fig:qualitative_75}]{%
        \scalebox{0.99}{\includegraphics[%
            width=0.155\textwidth,
            trim = {0 0 0 15},
            clip
        ]{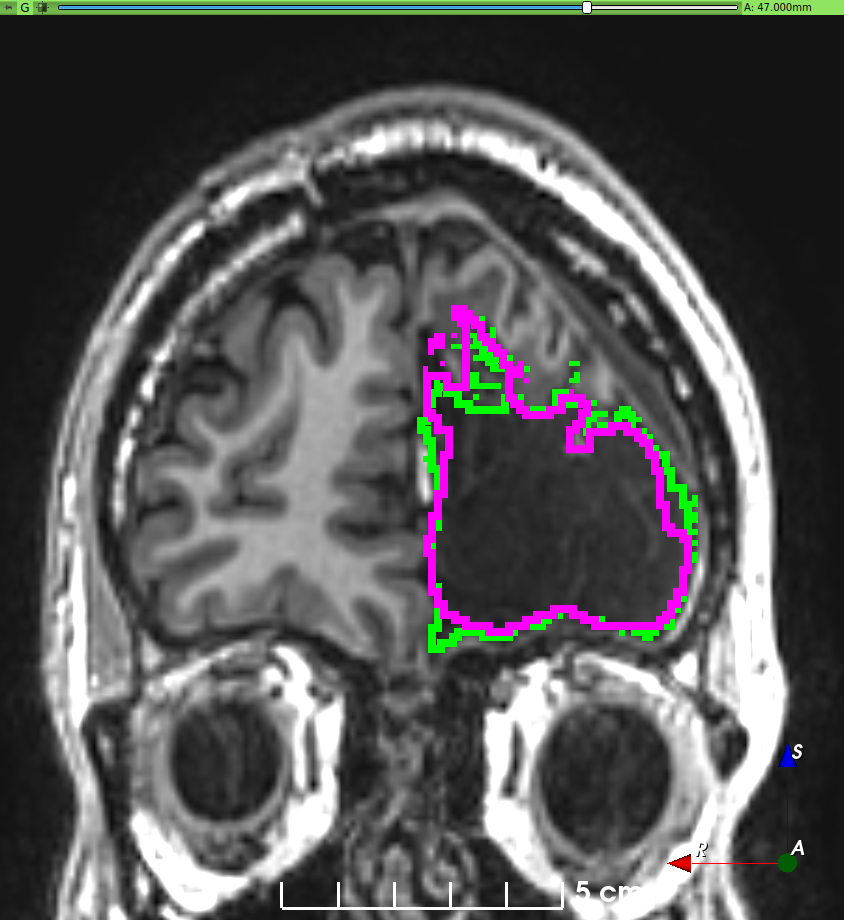}}
    }%
    \subfloat[\label{fig:qualitative_100}]{%
        \scalebox{0.99}{\includegraphics[%
            width=0.155\textwidth,
            trim = {0 0 0 15},
            clip
        ]{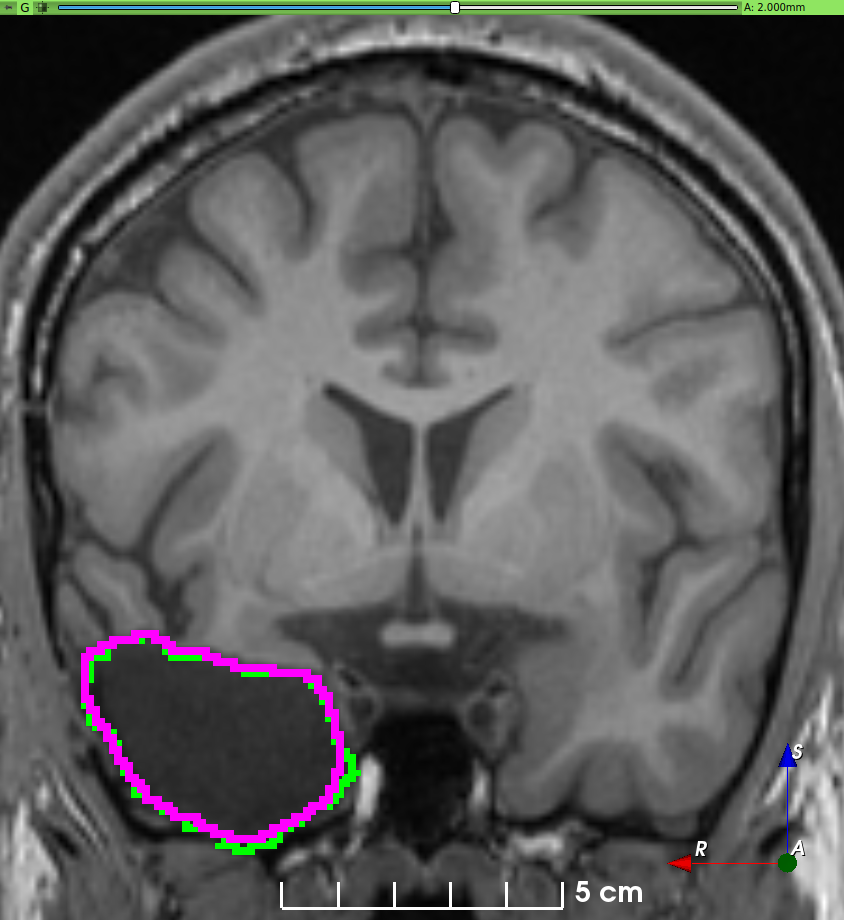}}
    }

    \caption{%
        Manual labels from Rater A (green) and
        Rater D, the model trained with \textit{PubFull} + \textit{EpiPre} + \textit{Pseudo} (magenta).
        Errors caused by a
        (a) small resection,
        (b) blood clot in cavity and
        (c) brain shift;
        segmentations corresponding to the
        (d) 50th, (e) 75th and (f) 100th percentiles
        giving a \gls{dsc} of 81.7, 86.5 and 93.8, respectively.
    }
    \label{fig:qualitative}
\end{figure}

We trained and tested on the 133 images annotated by
Rater A, using 10-fold cross-validation, obtaining a \gls{dsc} of 65.3 (30.6).
We refer to this dataset as \textit{EpiPost}.
For all other models,
we use data without manual annotations for training and
\textit{EpiPost} for testing.

\textit{EpiPre} comprised 261 preoperative MR images from
patients scanned at \gls{nhnn}
who underwent epilepsy surgery but are not in \textit{EpiPost}.
The model trained with \textit{EpiPre}
gave a \gls{dsc} of 61.6 (36.6), which was
not significantly different compared to
training with \textit{EpiPost} ($p = 0.216$).



We trained a model using \textit{PubSmall}, i.e.\ 261
images randomly chosen from the publicly available datasets.
This model had a \gls{dsc} of 69.5 (27.0).

Although there was a moderate increase in \gls{dsc},
training with either \textit{EpiPre} or \textit{PubSmall} was not significantly superior compared to \textit{EpiPost}
after Bonferroni correction ($p = 0.009$ and $p = 0.035$, respectively).

\subsection{Large Datasets}

We trained a model using the full public dataset (\textit{PubFull}, 1813 images),
obtaining a \gls{dsc} of 79.6 (17.3),
which was significantly superior to
\textit{PubSmall} ($p \approx 10^{-8}$) and \textit{EpiPost} ($p \approx 10^{-13}$).
%
Adding \textit{EpiPre} to \textit{PubFull}
for training did not significantly increase
performance (${p = 0.173}$),
with a \gls{dsc} of 80.5 (16.1).

For an additional training dataset,
we created the \textit{PubMed} dataset
by replacing 261 images in \textit{PubFull} with \textit{EpiPre}.
Training with \textit{PubMed} + \textit{EpiPre} was not significantly different compared to training with \textit{PubFull} ($p = 0.378$),
with a \gls{dsc} of 79.8 (17.1).

\subsection{Semi-supervised Learning}

We evaluated the ability of semi-supervised learning to improve model performance by generating pseudo-labels for all unlabeled postoperative images in EPISURG (297).
Pseudo-labels were generated by inferring the resection cavity label using the model trained on \textit{PubFull} and \textit{EpiPre}.
The pseudo-labels and corresponding postoperative images were combined to create the \textit{Pseudo} dataset.

We trained a model using
\textit{PubFull}, \textit{EpiPre} and \textit{Pseudo} (2371 images), obtaining a \gls{dsc} of 81.7 (14.2).
Adding the pseudo-labels to \textit{PubFull} and \textit{EpiPre} did not significantly improve performance ($p = 0.176$), indicating our semi-supervised learning approach provided no advantage.
Predictions from this model are shown in \cref{fig:qualitative}.

\subsection{Comparison to Inter-Rater Performance}

We computed pairwise inter-rater agreement between the three human raters
and the best performing model
(trained with \textit{PubFull} + \textit{EpiPre} + \textit{Pseudo}) as Rater D.

\begin{figure}[t!]
    \includegraphics[%
        width=\textwidth,
        trim={0 10 0 10},
        clip
    ]{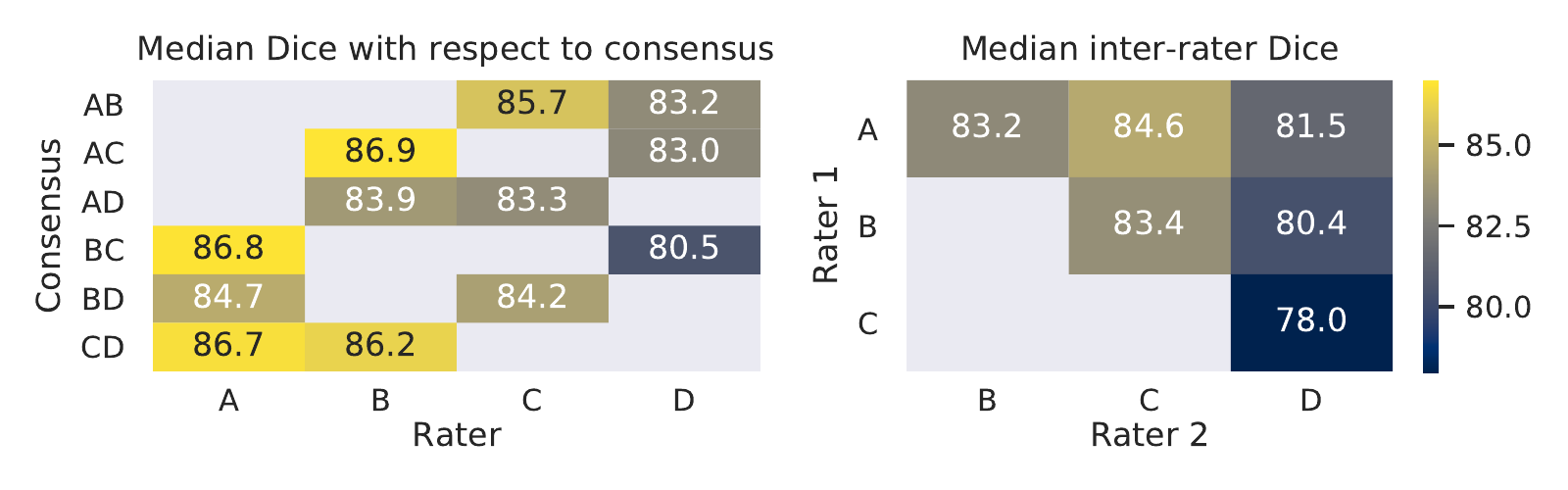}
    \caption{%
        Left: median \gls{dsc} between segmentations by a rater
        and consensuses from two other raters;
        right: median \gls{dsc} between each rater segmentations.
        Rater D corresponds to the model trained with \textit{PubFull}, \textit{EpiPre} and \textit{Pseudo}.
    }
    \label{fig:agreements}
\end{figure}

We computed consensus annotations between all pairs of raters
using shape-based averaging~\cite{rohlfing_shape-based_2007}.
\glspl{dsc} between the segmentations from each rater and
the consensuses generated by the other raters are
reported in \cref{fig:agreements}.

%% file: sections/discussion.tex
\section{Discussion}

We developed a method to simulate resection cavities on preoperative
\linebreak
$T_1$-weighted MR images and performed extensive validation using datasets
of different provenance and size.
Our results demonstrate that,
when the dataset is of a sufficient size,
simulating resection from unlabeled data
can provide more accurate segmentations
compared to a smaller manually annotated dataset.
%
%
We found that the most important factor for \gls{cnn} performance is
using a training dataset of sufficient size (in this example, 1800+ samples).
The inclusion of training samples from the same scanner or with pseudo-labels
only marginally improved the model performance.
However, we did not post-process the automatically-generated pseudo-labels,
nor did we exclude predictions with higher uncertainty.
Further improvements may be obtained by using more advanced semi-supervised
learning techniques to appropriately select pseudo-labels to use for training.

Predictions errors are mostly due to
1) resection of size comparable to sulci (\cref{fig:qualitative_small}),
2) unanticipated intensities, such as those caused by the presence of blood
clots in the cavity (\cref{fig:qualitative_texture}),
3) brain shift (\cref{fig:qualitative_brain_shift}) and
4) white matter hypointensities (\cref{fig:qualitative_75}).
%
%
Further work will involve
using different internal and external cavity textures,
carefully sampling the resection volume,
simulating brain shift using biomechanical models, and
quantifying epistemic and aleatoric segmentation uncertainty to better assess
model performance~\cite{shaw_heteroscedastic_2020}.

The model has a lower inter-rater agreement score
compared to between-human agreement values,
however, this is well within the interquartile range of all the
agreement values computed (\cref{fig:agreements}).
EPISURG will be made available,
so that it may be used as a benchmark dataset for brain cavity segmentation.

%% file: sections/acknowledgments.tex
\section*{Acknowledgments}

The authors wish to thank Luis García-Peraza Herrera and Reuben Dorent for the
fruitful discussions.

This work is supported by the UCL EPSRC Centre for Doctoral Training in
Medical Imaging (EP/L016478/1).
This publication represents in part independent research commissioned by the
Wellcome Trust Health Innovation Challenge Fund (WT106882).
The views expressed in this publication are those of the authors and not
necessarily those of the Wellcome Trust.

This work uses data provided by patients and collected by the \gls{nhs} as part
of their care and support.

%% file: sections/supplementary.tex
\section*{Supplementary Material}

\begin{table}
    \centering  
    \caption{%
        Datasets used for training. \gls{nhnn} refers to \acrlong{nhnn}. See
        \cref{sec:data} for more information about data provenance.
    }
    \label{tab:datasets}
    \begin{tabular}{*5c}
        \toprule
        \textbf{Name} & \textbf{Subjects} & \textbf{Source} & \textbf{Type} & \textbf{Annotations} \\
        \midrule
        \textit{EpiPost}  &  133 & \gls{nhnn} & Postoperative & Yes \\
        \textit{EpiPre}   &  261 & \gls{nhnn} & Preoperative  & No  \\
        \textit{Pseudo}   &  297 & \gls{nhnn} & Postoperative & No  \\
        \textit{PubSmall} &  261 & Public     & Preoperative  & No  \\
        \textit{PubMed}   & 1552 & Public     & Preoperative  & No  \\
        \textit{PubFull}  & 1813 & Public     & Preoperative  & No  \\
        \bottomrule
    \end{tabular}
\end{table}

\begin{table}
    \centering  
    \caption{%
        \gls{dsc} values between manual annotations from Rater A
        and models segmentations, as shown in \cref{fig:boxes}.
        Note that only the first model was trained with manual annotations.
        \textit{EpiPost}: postoperative scans in EPISURG with manual annotations (*~indicates fully-supervised training with 10-fold cross-validation);
        \textit{EpiPre}: preoperative scans from subjects not contained in \textit{EpiPost});
        \textit{PubFull}: public datasets;
        \textit{PubSmall}, \textit{PubMed}: subsets of \textit{PubFull};
        \textit{Pseudo}: pseudo-labeled postoperative scans in EPISURG.
    }
    \label{tab:quantitative}
    \begin{tabular}{*4c}
        \toprule
        \textbf{Training dataset} & \textbf{Subjects} & \textbf{Annotations} & \textbf{Dice score} \\
        \midrule
        \textit{EpiPost}*                                    &  133 & Yes &          65.3 (30.6)   \\
        \textit{EpiPre}                                      &  261 & No  &          61.6 (36.6)   \\
        \textit{PubSmall}                                    &  261 & No  &          69.5 (27.0)   \\
        \textit{PubFull}                                     & 1813 & No  &          79.6 (17.3)   \\
        \textit{PubMed}  + \textit{EpiPre}                   & 1813 & No  &          79.8 (17.1)   \\
        \textit{PubFull} + \textit{EpiPre}                   & 2074 & No  &          80.5 (16.1)   \\
        \textit{PubFull} + \textit{EpiPre} + \textit{Pseudo} & 2371 & No  &  \textbf{81.7 (14.2)}  \\
        \bottomrule
    \end{tabular}
\end{table}

\begin{figure}[b!]
    \centering

    \captionsetup[subfloat]{farskip=2pt,captionskip=1pt}

    \subfloat{%
        \scalebox{0.99}{\includegraphics[%
            width=0.13\textwidth,
        ]{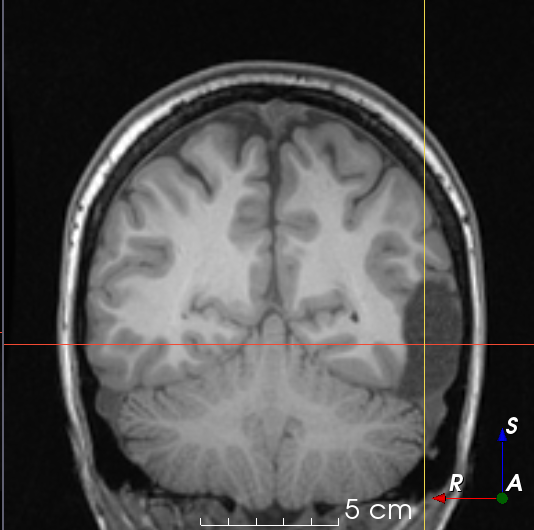}}
    }%
    \subfloat{%
        \scalebox{0.99}{\includegraphics[%
            width=0.13\textwidth,
        ]{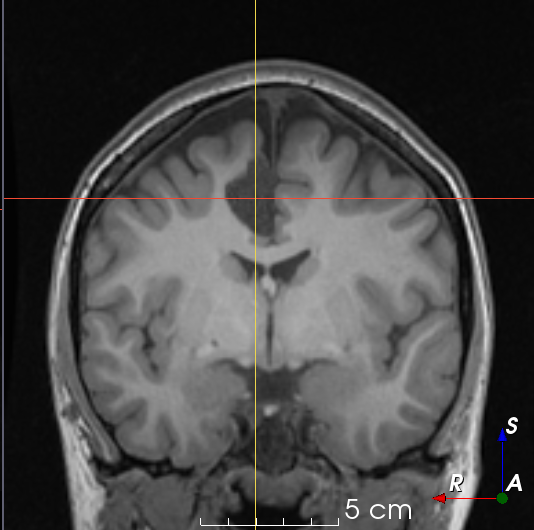}}
    }%
    \subfloat{%
        \scalebox{0.99}{\includegraphics[%
            width=0.13\textwidth,
        ]{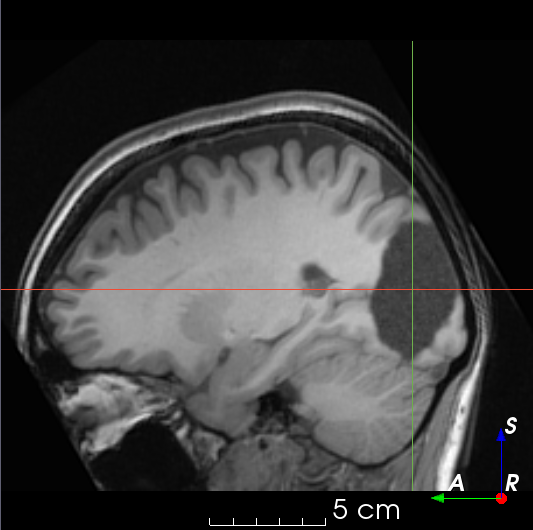}}
    }%
    \subfloat{%
        \scalebox{0.99}{\includegraphics[%
            width=0.13\textwidth,
        ]{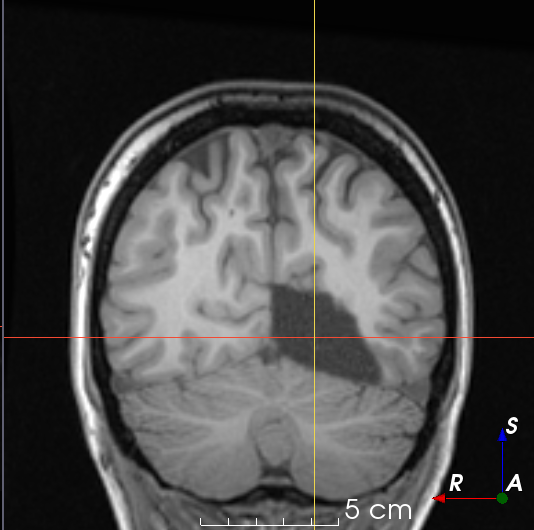}}
    }%
    \subfloat{%
        \scalebox{0.99}{\includegraphics[%
            width=0.13\textwidth,
        ]{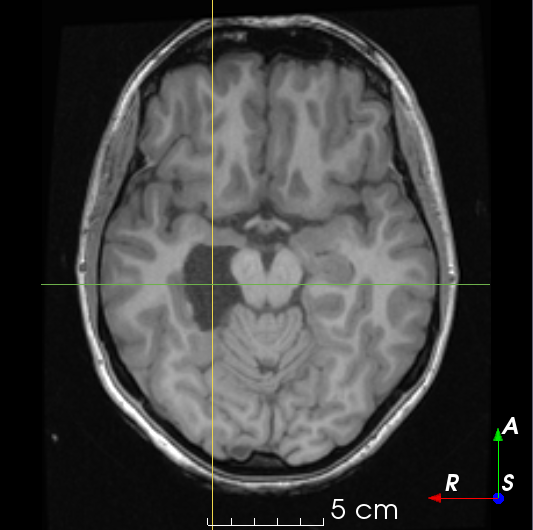}}
    }
    
    \subfloat{%
        \scalebox{0.99}{\includegraphics[%
            width=0.13\textwidth,
        ]{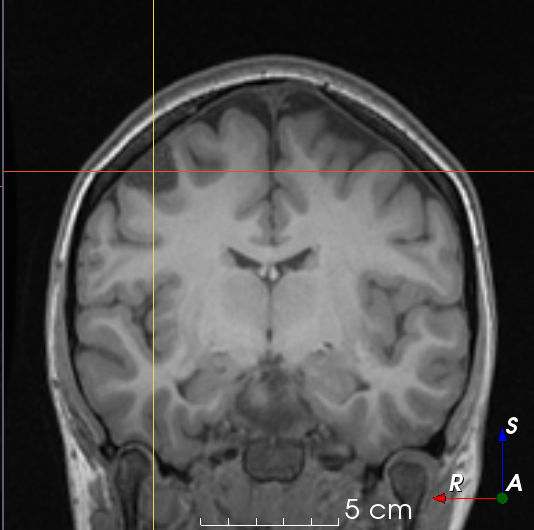}}
    }%
    \subfloat{%
        \scalebox{0.99}{\includegraphics[%
            width=0.13\textwidth,
        ]{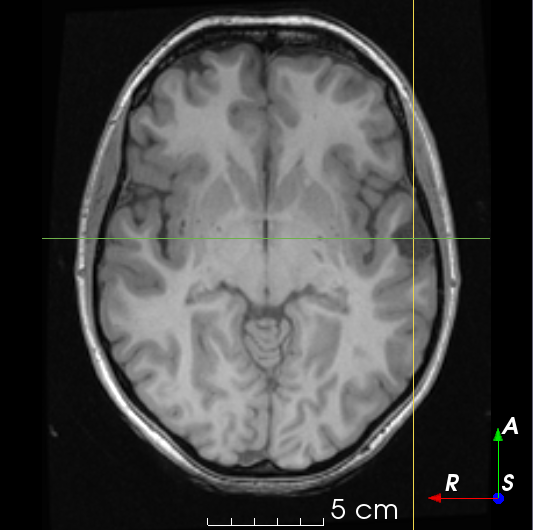}}
    }%
    \subfloat{%
        \scalebox{0.99}{\includegraphics[%
            width=0.13\textwidth,
        ]{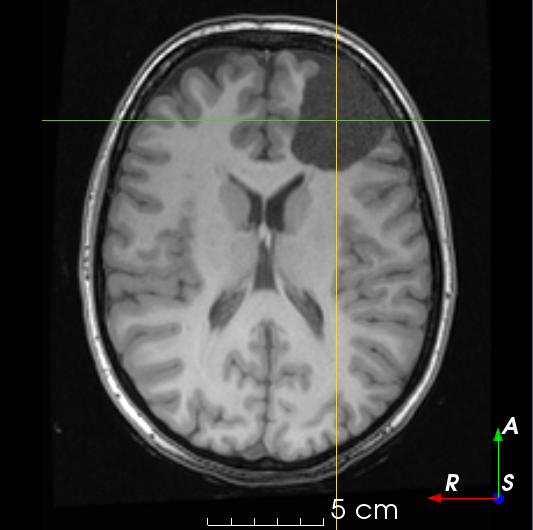}}
    }%
    \subfloat{%
        \scalebox{0.99}{\includegraphics[%
            width=0.13\textwidth,
        ]{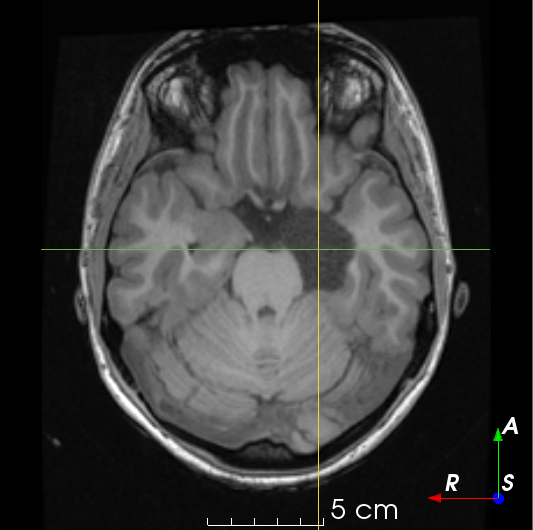}}
    }%
    \subfloat{%
        \scalebox{0.99}{\includegraphics[%
            width=0.13\textwidth,
        ]{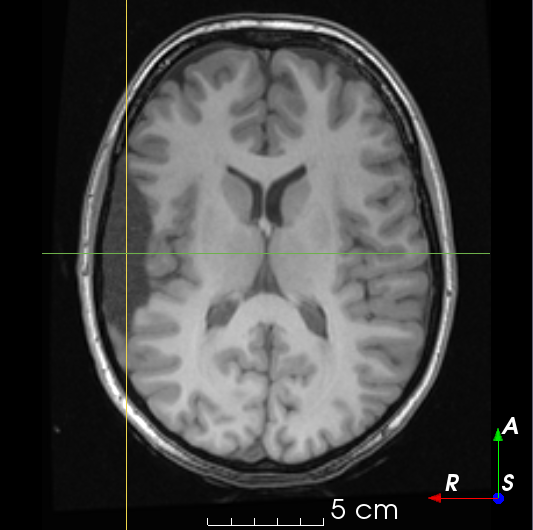}}
    }
    
    \caption{%
        Resection simulations $\X_R$ generated using our method.
    }
    \label{fig:simulations}
\end{figure}

\begin{figure}
    \centering

    \subfloat[\label{fig:qualitative_small}]{%
        \scalebox{0.99}{\includegraphics[%
            width=0.5\textwidth,
            trim = {0 0 0 50},
            clip
        ]{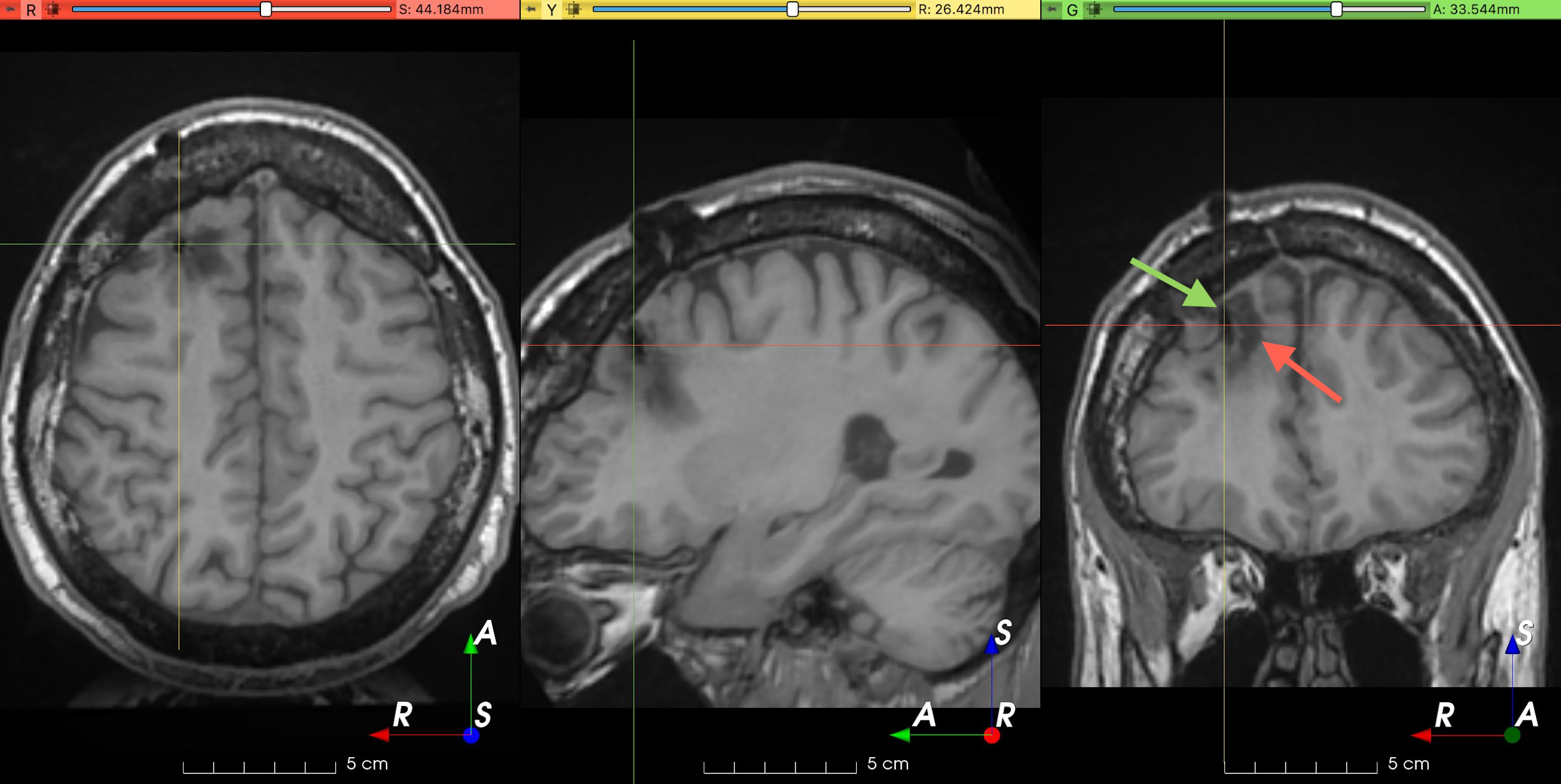}}
    }%
    \subfloat[\label{fig:qualitative_texture}]{%
        \scalebox{0.99}{\includegraphics[%
            width=0.5\textwidth,
            trim = {0 0 0 50},
            clip
        ]{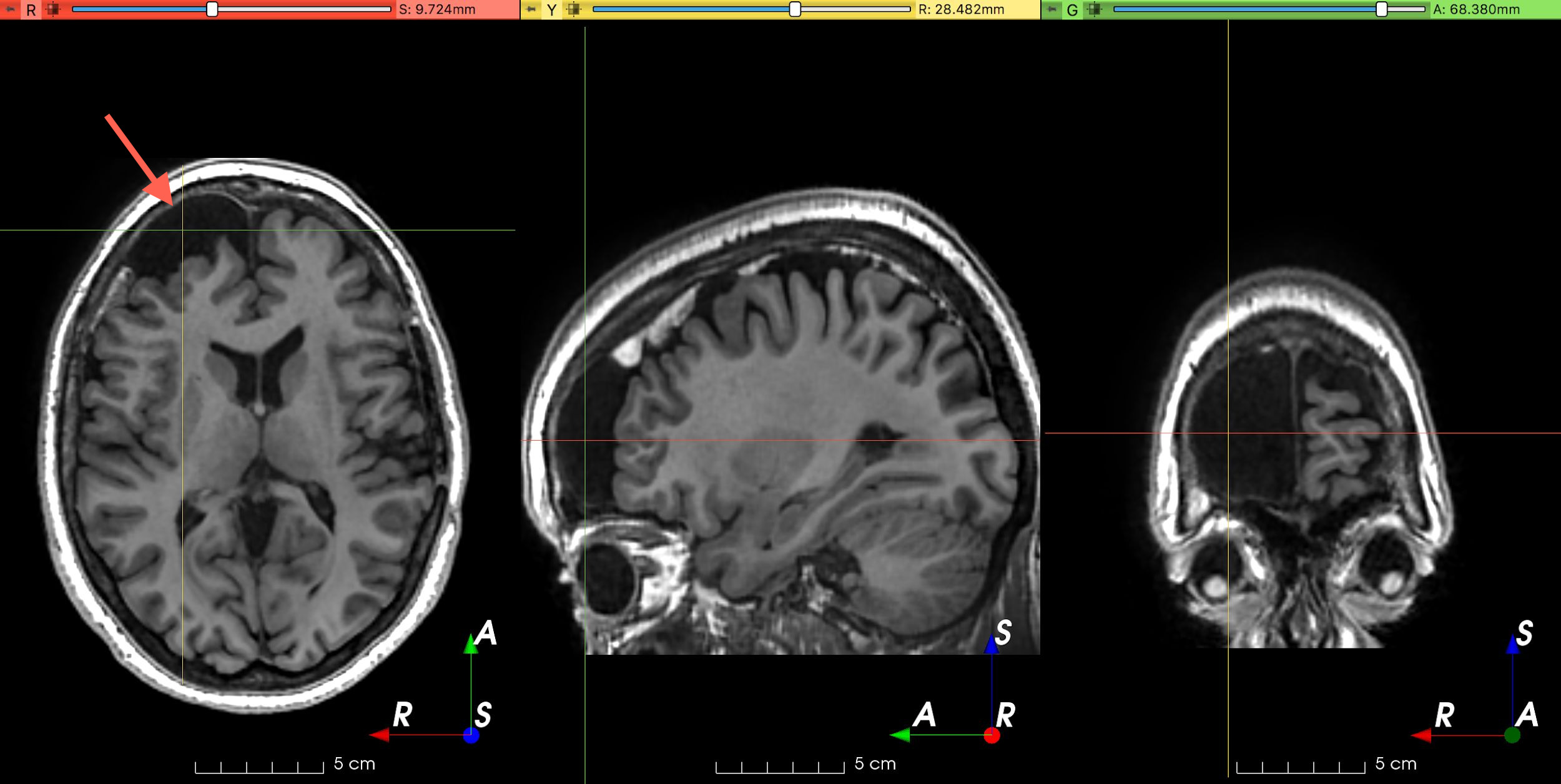}}
    }

    \subfloat[\label{fig:qualitative_brain_shift}]{%
        \scalebox{0.99}{\includegraphics[%
            width=0.5\textwidth,
            trim = {0 0 0 50},
            clip
        ]{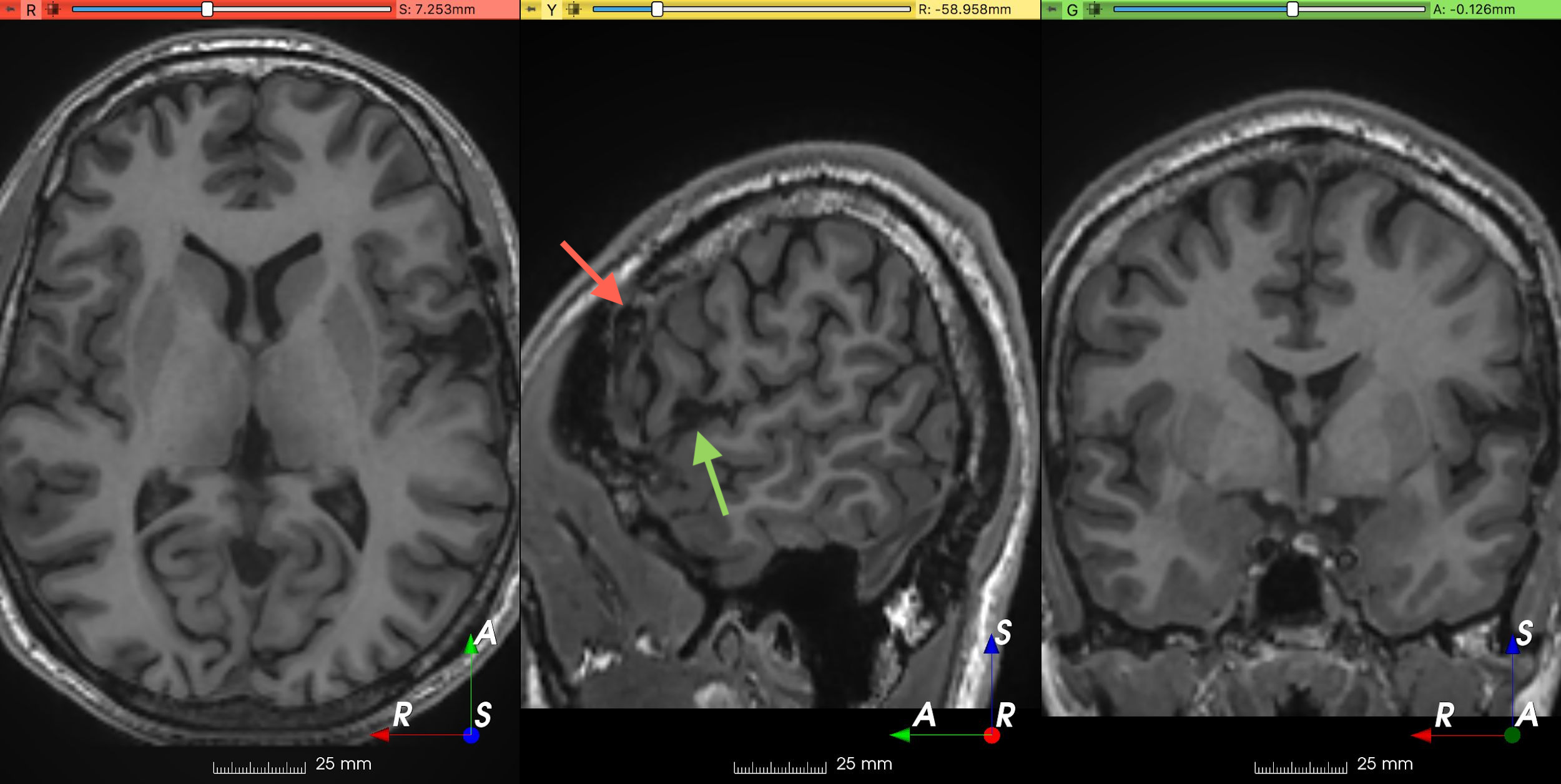}}
    }
    \subfloat[\label{fig:qualitative_50}]{%
        \scalebox{0.99}{\includegraphics[%
            width=0.5\textwidth,
            trim = {0 0 0 50},
            clip
        ]{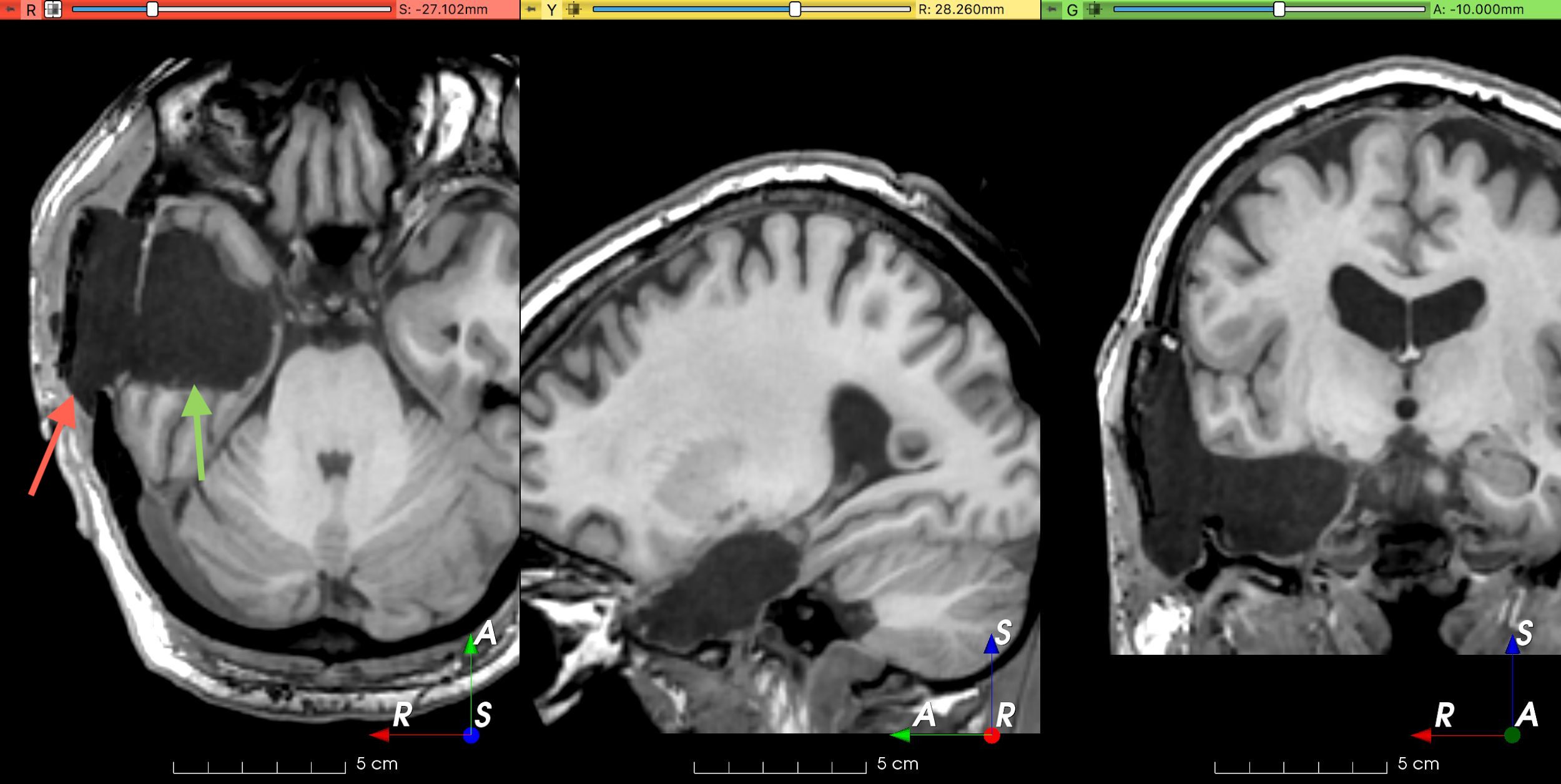}}
    }

    \subfloat[\label{fig:qualitative_75}]{%
        \scalebox{0.99}{\includegraphics[%
            width=0.5\textwidth,
            trim = {0 0 0 50},
            clip
        ]{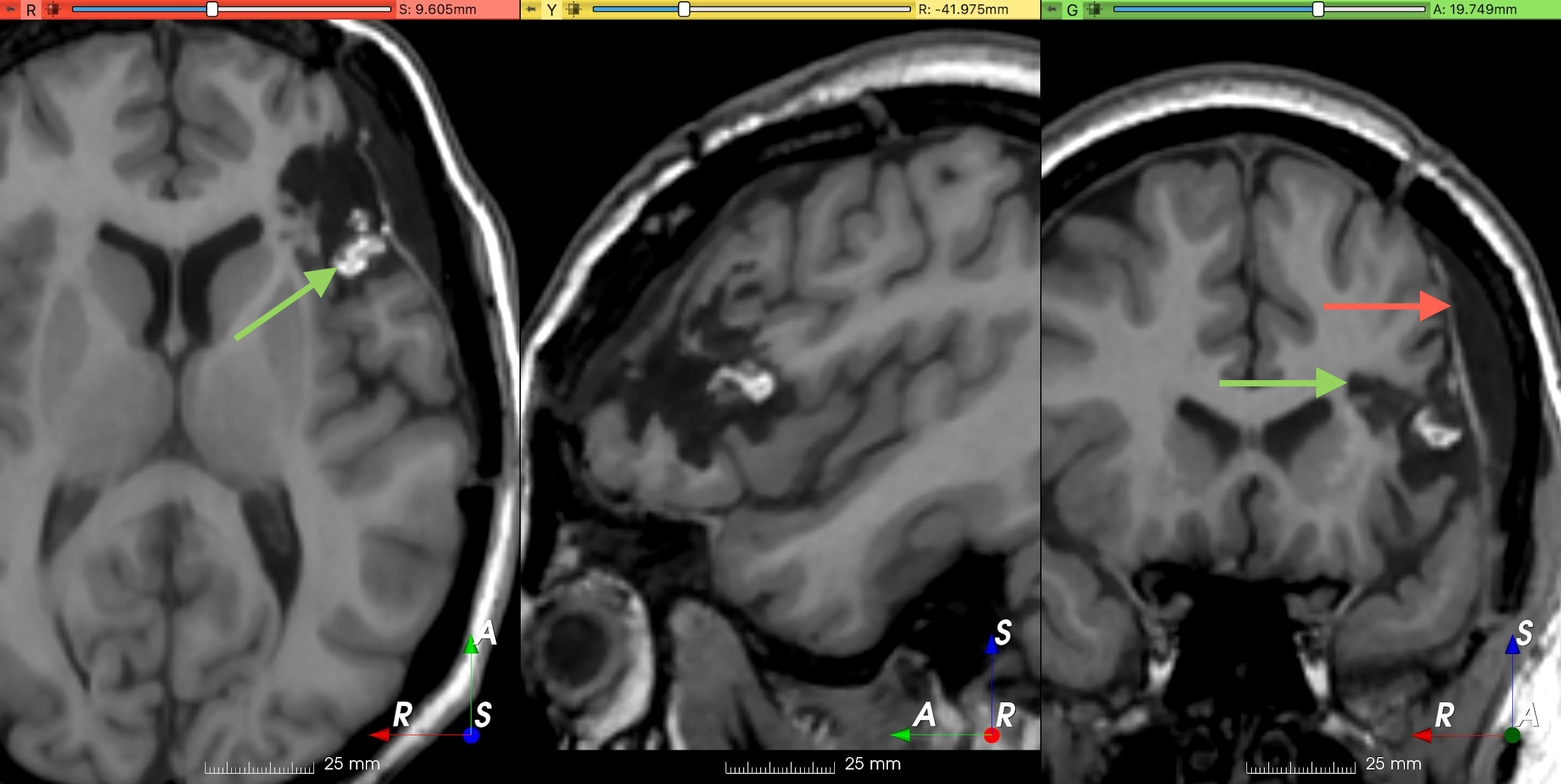}}
    }%
    \subfloat[\label{fig:qualitative_100}]{%
        \scalebox{0.99}{\includegraphics[%
            width=0.5\textwidth,
            trim = {0 0 0 50},
            clip
        ]{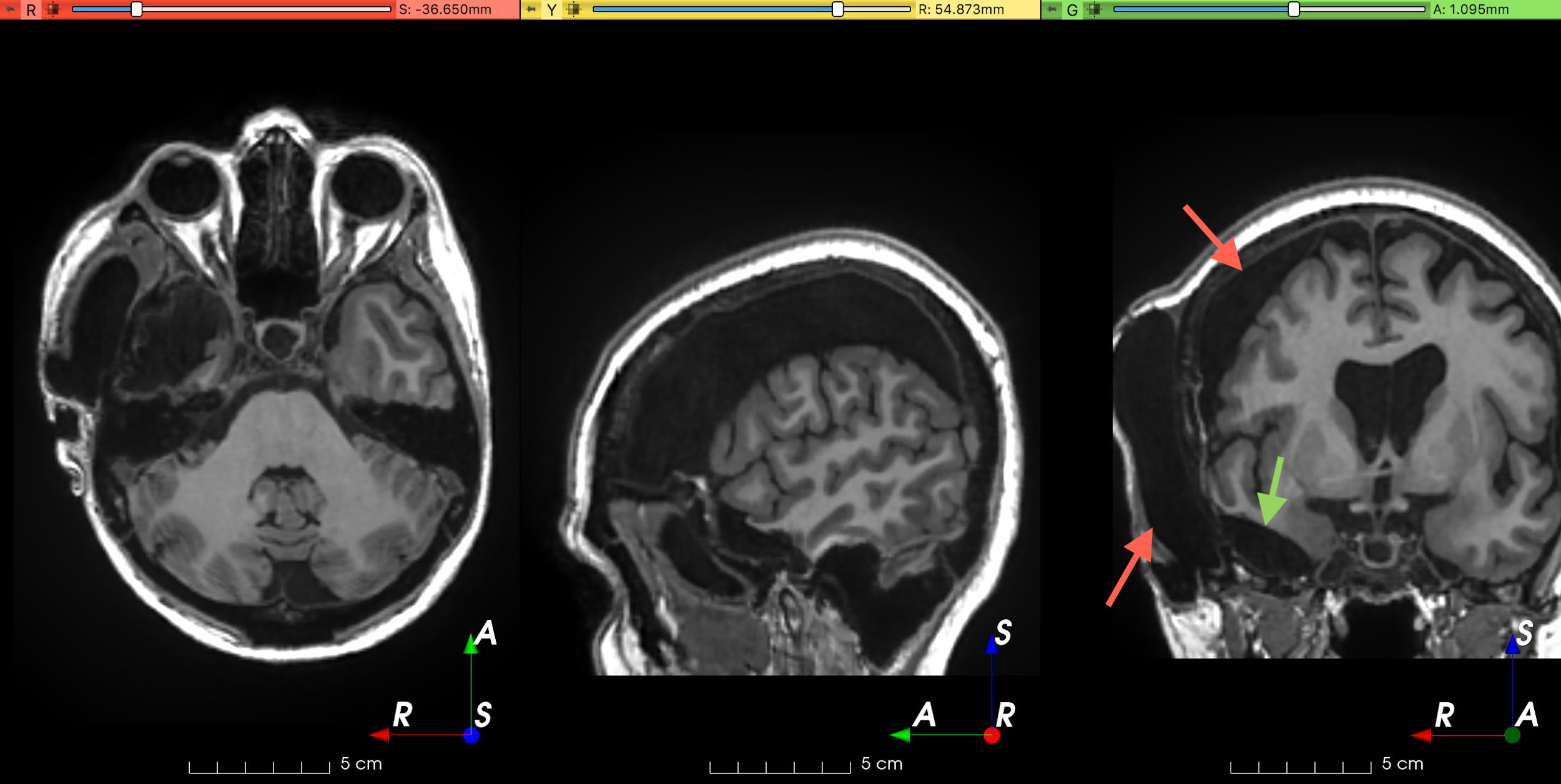}}
    }

    \subfloat[\label{fig:qualitative_100}]{%
        \scalebox{0.99}{\includegraphics[%
            width=0.5\textwidth,
            trim = {0 0 0 50},
            clip
        ]{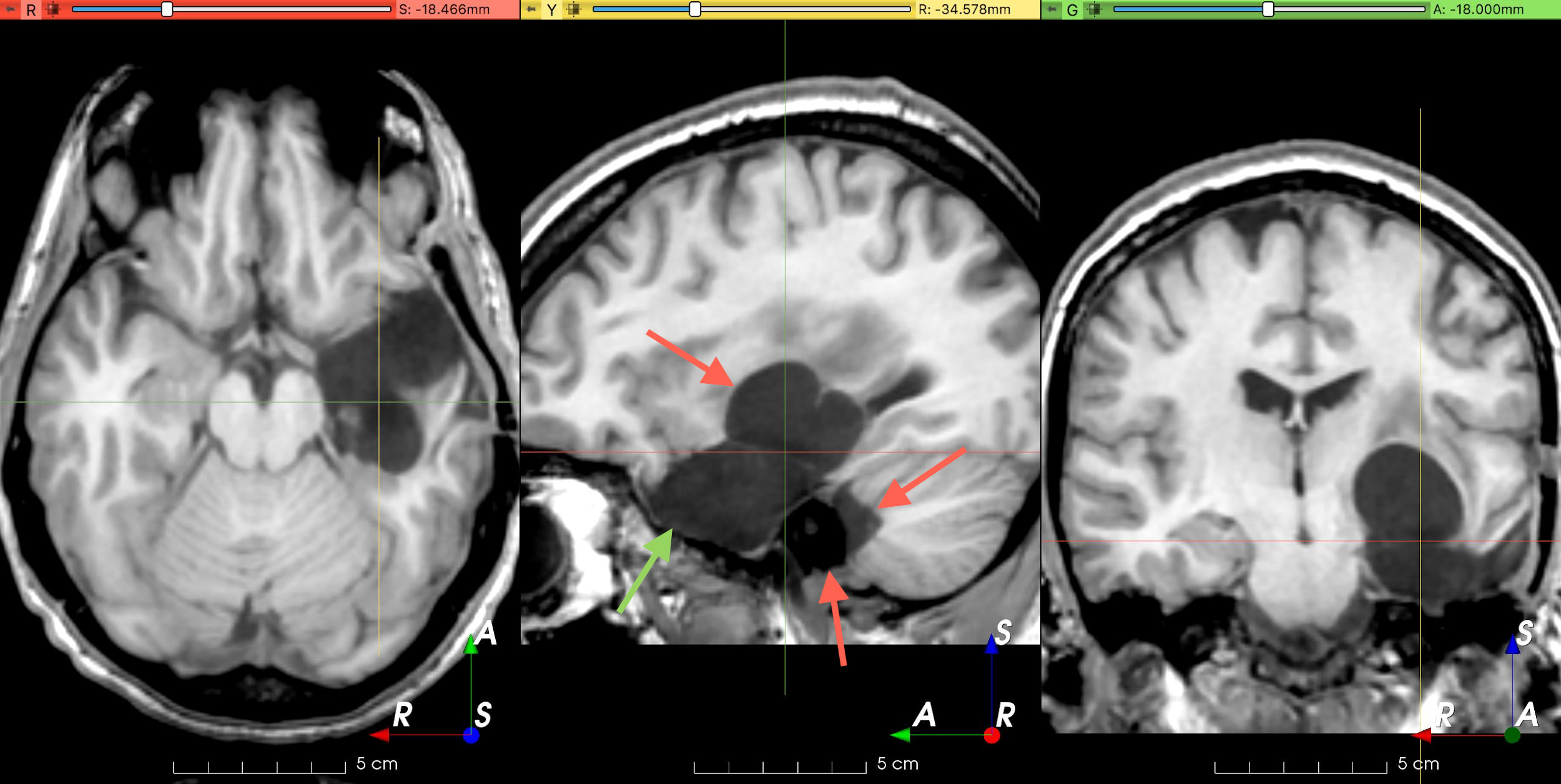}}
    }
    \subfloat[\label{fig:qualitative_100}]{%
        \scalebox{0.99}{\includegraphics[%
            width=0.5\textwidth,
            trim = {0 0 0 50},
            clip
        ]{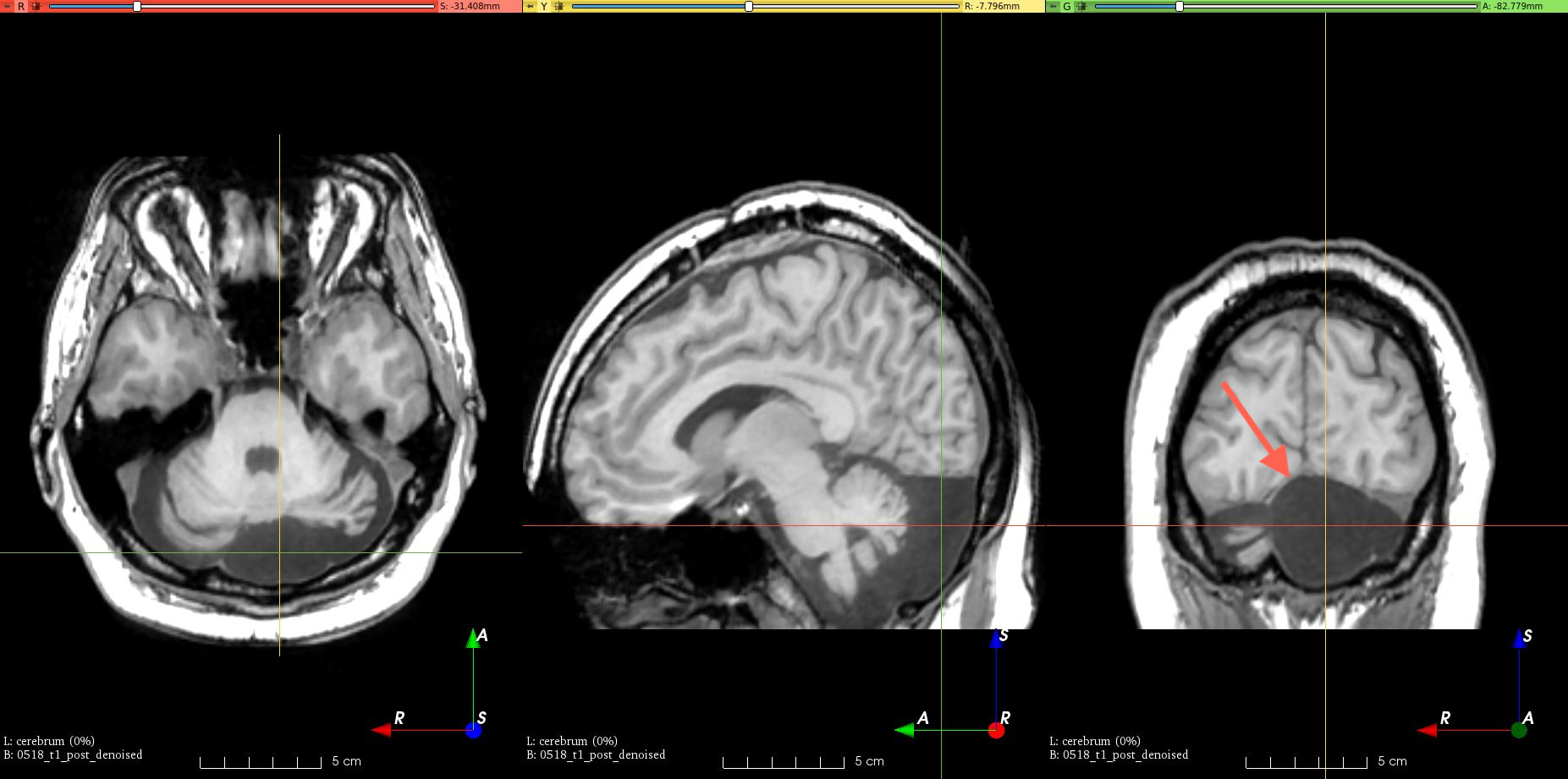}}
    }

    \caption{%
        Examples of challenging images for cavity segmentation.
        (a)~Small frontal lesionectomy surrounded by hypointense white matter
        (b)~Brain shift after contralateral temporal lobectomy (not shown)
        (c)~Small frontal lesionectomy near the Sylvian fissure
        (d)~Lack of boundaries between oedema and resection cavity
        (e)~Possible blood clot within the cavity
        (f)~Brain shift, oedema and resection cavity
        (g)~Arachnoid cyst and resection cavity
        (h)~Cerebellar degeneration.
        Green annotations represent areas that correspond to resection cavities; red annotations represent areas that do not.
    }
    \label{fig:challenging}
\end{figure}